\documentclass[twocolumn]{aastex631}
\usepackage{comment}

\newcommand\pRT{\texttt{petitRADTRANS}}

\defcitealias{lothringer:2022}{Lothringer \& Sing, et al., 2022} 

\received{January 31, 2025}
\revised{March 12, 2025}
\accepted{March 13, 2025}

\submitjournal{AJ}

\begin{document}

\title{Refractory and Volatile Species in the UV-to-IR Transmission Spectrum of Ultra-hot Jupiter WASP-178b with HST and JWST}

\author[0000-0003-3667-8633]{Joshua D. Lothringer}
\affiliation{Space Telescope Science Institute, Baltimore, MD}

\author[0000-0002-9030-0132]{Katherine A. Bennett}
\affiliation{Department of Earth and Planetary Sciences, Johns Hopkins University, Baltimore, MD}

\author[0000-0001-6050-7645]{David K. Sing}
\affiliation{Department of Earth and Planetary Sciences, Johns Hopkins University, Baltimore, MD}
\affiliation{William H.\ Miller III Department of Physics \& Astronomy,
Johns Hopkins University, 3400 N Charles St, Baltimore, MD 21218, USA}

\author{Brian Kehoe-Seamons}
\affiliation{Department of Physics, Utah Valley University, Orem, UT}

\author[0000-0003-4408-0463]{Zafar Rustamkulov}
\affiliation{Department of Earth and Planetary Sciences, Johns Hopkins University, Baltimore, MD}

\author[0000-0001-6533-6179]{Henrique Reggiani}
\affiliation{Gemini South, Gemini Observatory, NSF's NOIRLab, Casilla 603,
La Serena, Chile}

\author[0000-0001-5761-6779]{Kevin C.\ Schlaufman}
\affiliation{William H.\ Miller III Department of Physics \& Astronomy,
Johns Hopkins University, 3400 N Charles St, Baltimore, MD 21218, USA}

\author[0000-0003-0473-6931]{Patrick McCreery}
\affiliation{William H.\ Miller III Department of Physics \& Astronomy,
Johns Hopkins University, 3400 N Charles St, Baltimore, MD 21218, USA}

\author{Seti Norris}
\affiliation{Space Telescope Science Institute, Baltimore, MD}

\author[0000-0002-8077-5572]{Peter Hauschildt}
\affiliation{Hamburger Sternwarte, Gojenbergsweg 112, 21029 Hamburg, Germany}

\author{Ceiligh Cacho-Negrete}
\affiliation{Space Telescope Science Institute, Baltimore, MD}

\author[0000-0003-0854-3002]{Am\'elie Gressier}
\affiliation{Space Telescope Science Institute, Baltimore, MD}

\author[0000-0001-9513-1449]{N\'estor Espinoza}
\affiliation{Space Telescope Science Institute, Baltimore, MD}

\author[0009-0007-9356-8576]{Cyril Gapp}
\affiliation{Max-Planck-Institut f\"{u}r Astronomie, K\"{o}nigstuhl 17, D-69117 Heidelberg, Germany}
\affiliation{Department of Physics and Astronomy, Heidelberg University, Im Neuenheimer Feld 226, D-69120 Heidelberg, Germany}

\author[0000-0001-5442-1300]{Thomas M. Evans-Soma}
\affiliation{School of Information and Physical Sciences, University of Newcastle, Callaghan, NSW, Australia}
\affiliation{Max-Planck-Institut f\"{u}r Astronomie, K\"{o}nigstuhl 17, D-69117 Heidelberg, Germany}

\author[0000-0002-7352-7941]{Kevin B.\ Stevenson}
\affiliation{Johns Hopkins APL, 11100 Johns Hopkins Rd, Laurel, MD 20723, USA}

\author[0000-0003-4328-3867]{Hannah Wakeford}
\affiliation{School of Physics, University of Bristol, H.H. Wills Physics Laboratory, Bristol, UK}

\author[0000-0002-9308-2353]{Neale Gibson}
\affiliation{Trinity College Dublin, University of Dublin, Dublin, Ireland}

\author{Jamie Wilson}
\affiliation{Department of Mathematics and Statistics, University of Limerick, Limerick, Ireland}

\author[0000-0002-6500-3574]{Nikolay Nikolov}
\affiliation{Space Telescope Science Institute, Baltimore, MD}


\begin{abstract}

\noindent
The atmospheres of ultra-hot Jupiters are unique compared to other planets because of the presence of both refractory and volatile gaseous species, enabling a new lens to constrain a planet's composition, chemistry, and formation. WASP-178b is one such ultra-hot Jupiter that was recently found to exhibit enormous NUV absorption between 0.2 and 0.4 $\mu$m from some combination of Fe+, Mg, and SiO. Here, we present new infrared observations of WASP-178b with HST/WFC3 and JWST/NIRSpec/G395H, providing novel measurements of the volatile species H$_2$O and CO in WASP-178b's atmosphere. Atmospheric retrievals find a range of compositional interpretations depending on which dataset is retrieved, the type of chemistry assumed, and the temperature structure parametrization used due to the combined effects of thermal dissociation, the lack of volatile spectral features besides H$_2$O and CO, and the relative weakness of H$_2$O and CO themselves. Taken together with a new state-of-the-art characterization of the host star, our retrieval analyses suggests a solar to super-solar [O/H] and [Si/H], but sub-solar [C/H], perhaps suggesting rock-laden atmospheric enrichment near the H$_2$O iceline. To obtain meaningful abundance constraints for this planet, it was essential to combine the JWST IR data with short-wavelength HST observations, highlighting the ongoing synergy between the two facilities. 

\end{abstract}


\section{Introduction} \label{sec:intro}

Ultra-hot Jupiters are roughly defined as planets above about 2000--2200~K. Around this temperature, theoretical studies have suggested a fundamental transition exists from cloudy, neutral, molecule-dominated atmospheres to cloud-free, ion-rich, atom-dominated stratospheres \citep{lothringer:2018b,parmentier:2018,kitzmann:2018}. So far, these predictions generally appear to have been borne out by a wide number of observations of such systems \citep[e.g.,][]{mansfield:2020b,mansfield:2020,baxter:2020,gao:2020,hoeijmakers:2018a}.

One of the most unique aspects of ultra-hot Jupiters is the ability to measure refractory elements in their atmospheres, thanks to the fact that their atmospheres remain hot enough to avoid the condensation of such species, at least on their dayside. These refractory elements have been seen through a variety of high-resolution observations, both in transmission and emission \citep[e.g.,][]{hoeijmakers:2018a,hoeijmakers:2019}, while also recently being detected in the gaseous molecular form of SiO in the JWST/NIRSpec/G395H phase curve of WASP-121b \citep{evans-soma:2025,gapp:2025}.

Ideally, this variety of species can be used to measure the bulk refractory and volatile ratios that one can then use to infer a rock-to-ice ratio. Such a measurement is extremely useful because it probes two different reservoirs of material available during planet formation and can therefore be used to more precisely constrain the formation mechanism and migration history of a planet \citep[][]{turrini:2021,schneider:2021,lothringer:2021,pancetti:2022,chachan:2023}. This is especially critical since the C/O ratio alone is not thought to be enough information to uniquely constrain such formation histories, even as simply as identifying whether a planet formed inside or outside the H$_2$O ice line \citep[e.g.,][]{mordasini:2016}. 

\subsection{The WASP-178 System}

WASP-178b/KELT-26b (hereafter WASP-178b) is a $T_\mathrm{eq}=2470\pm60$~K ultra-hot Jupiter orbiting the $V=9.95$ chemically peculiar A-star HD 134004, hereafter WASP-178 \citep{hellier:2019, rodriguezmartinez:2020}. The host star's high effective temperature at $9200^{+200}_{-170}$~K makes it the second-hottest known planet-hosting star, second only to KELT-9 \citep{gaudi:2017}. While this fact has consequences for the planet's atmosphere itself \citep[e.g.,][]{lothringer:2019}, it also means that WASP-178b is more amenable to characterization at short-wavelengths compared to nearly all other planets thanks to the bright UV ``backlight" from the host star.

To that end, HST/WFC3/G280 (0.2--0.8\,$\mu$m) observations of the system revealed an enormous NUV absorption feature in the planet's transmission spectrum extending nearly 20 scale-heights from the optical continuum that atmospheric retrievals identified as some combination of Mg, Fe, and SiO \citepalias{lothringer:2022}. This robustly implied that silicate clouds have not condensed at the terminator of WASP-178b, in line with other empirical trends \citep{gao:2020} and theoretical modeling \cite[e.g.,][]{visscher:2010}. Complementary high-resolution HST/STIS/E230M (0.23--0.31$\mu$m) observations were unable to detect escaping Mg+ or Fe+ from WASP-178b, suggesting that the atmosphere probed by NUV observations is still bound to the planet (though not necessarily hydrostatically).

Na, H$\alpha$, H$\beta$, Mg, Fe, and Fe+ were also recently detected in WASP-178b with high-resolution ESPRESSO transit observations \citep{damasceno:2024}, confirming the presence of these silicate cloud precursors at the terminator. Furthermore, \cite{pagano:2024} used CHEOPS and TESS to measure a dayside temperature of around 2250--2800~K and a geometric albedo between 0.1--0.35, also indicative of the absorption of short-wavelength light by refractory species, rather than strong reflection from clouds or hazes. Combining the CHEOPS and TESS eclipse observations with high-resolution CRIRES+ data, \cite{cont:2024} measured a super-solar metallicity and a solar C/O ratio on the dayside of WASP-178b.

Here, we present new transit observations with HST/WFC3/G102 and G141, extending the continuous short-wavelength coverage out to 1.7\,$\mu$m. We combine this with new JWST/NIRSpec/G395H observations from 2.8 to 5.1\,$\mu$m, providing a measurement of the carbon and oxygen content of WASP-178b's atmosphere to complement the refractory species detected at short wavelengths. We  first describe results of our characterization of the host star in Section~\ref{sec:stellar}. Section~\ref{sec:methods} describes the methods used to reduce and analyze the data planet's transmission spectrum, including atmospheric retrievals. Section~\ref{sec:results} explains the results of the data analysis and the atmospheric retrievals. Section~\ref{sec:discussion} provides a discussion of the results, including interpretations as to the planet's formation in Section~\ref{sec:formation}, followed by a conclusion in Section~\ref{sec:conclusion}.

\section{Stellar Characterization and System Properties}\label{sec:stellar}

To better understand the formation of the planet, we first refine our knowledge of the host star. We infer the fundamental and photospheric stellar parameters of WASP-178 using the \texttt{isochrones} \citep{mor15} package to execute with \texttt{MultiNest} \citep{fer08,fer09,fer19} a simultaneous Bayesian fit of the Modules for Experiments in Stellar Evolution \citep[MESA;][]{pax11,pax13,pax18,pax19,jer23} Isochrones \& Stellar Tracks \citep[MIST;][]{dot16,cho16} isochrone grid to a curated collection of data for the star.  We fit the MIST grid to

\begin{enumerate}
\item
SkyMapper Southern Survey DR4 $uvriz$ photometry including in
quadrature their zero-point uncertainties (0.03,0.02,0.01,0.01,0.02)
mag \citep{onk24};
\item
Gaia DR2 $G$ photometry including in quadrature its zero-point uncertainty
\citep{gai16,gaia:2018,are18,bus18,eva18,rie18};
\item
Two-micron All-sky Survey (2MASS) $JHK_{s}$ photometry including their
zero-point uncertainties \citep{skr06};
\item
Wide-field Infrared Survey Explorer (WISE) CatWISE2020 $W1W2$ photometry
including in quadrature their zero-point uncertainties (0.032,0.037)
mag\footnote{\url{https://wise2.ipac.caltech.edu/docs/release/allsky/expsup/sec4\_4h.html\#PhotometricZP}}
\citep{wri10,mai11,eis20,mar21};
\item
a zero point-corrected Gaia DR3 parallax
\citep{gai21,fab21,lin21a,lin21b,row21,tor21}; and
\item
an estimated extinction value based on a three-dimensional extinction
map \citep{lal22,ver22}.
\end{enumerate}
As priors we use
\begin{enumerate}
\item
a \citet{cha03} log-normal mass prior for $M_{\ast} < 1~M_{\odot}$
joined to a \citet{sal55} power-law prior for $M_{\ast} \geq 1~M_{\odot}$;
\item
a metallicity prior based on the Geneva-Copenhagen Survey
\citep[GCS;][]{cas11};
\item
a log-uniform age prior between 10 Myr and 10 Gyr;
\item
a uniform extinction prior in the interval 0 mag $< A_{V} < 0.5$ mag; and
\item
a distance prior proportional to volume between the \citet{bai21}
geometric distance minus/plus five times its uncertainty.
\end{enumerate}

\begin{figure}
    \centering
    \includegraphics[width=1\linewidth]{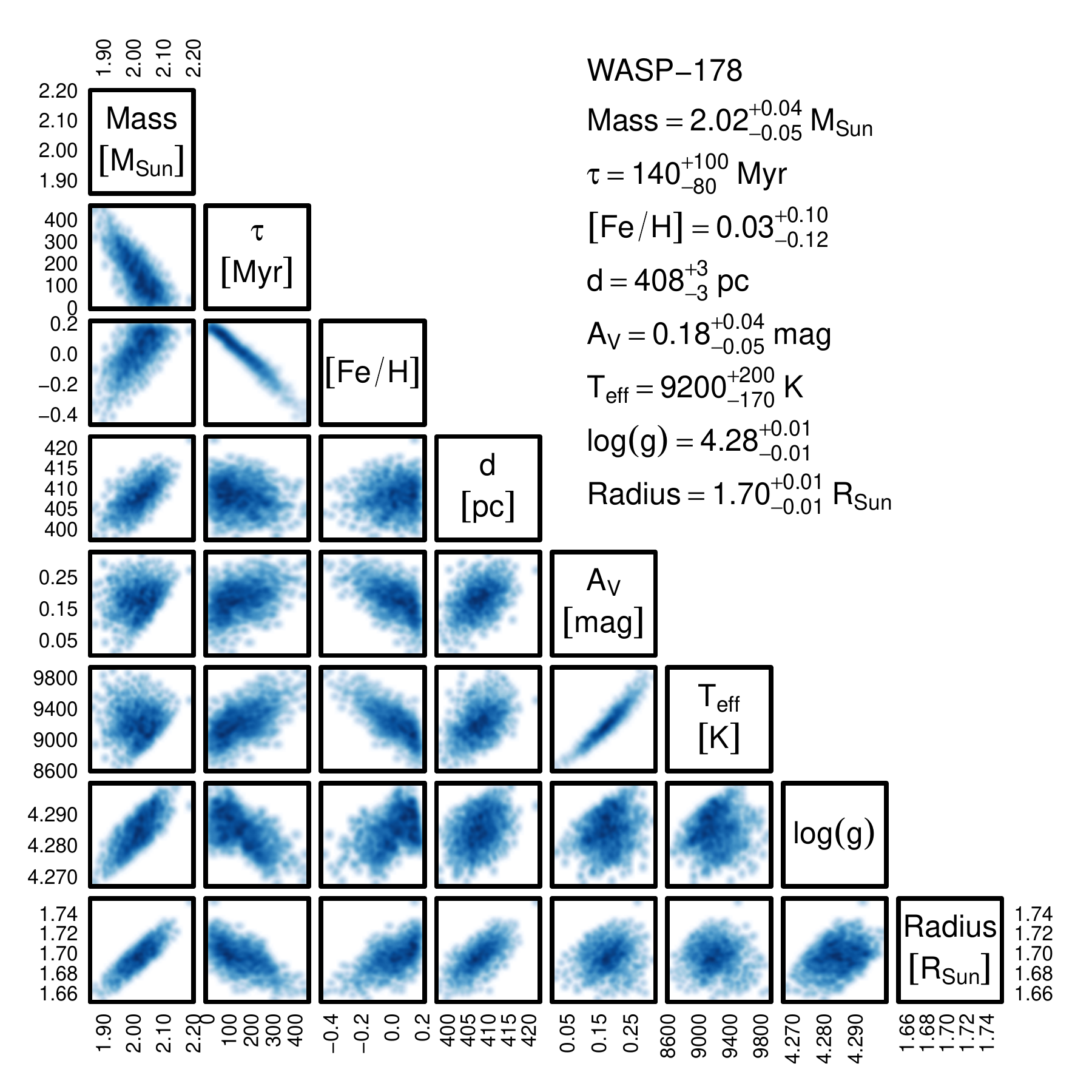}
    \caption{Posterior distribution from the stellar characterization in Section~\ref{sec:stellar}. Our analysis improves the precision on the fundamental stellar parameters, most significantly the age, which we constrain to be $140^{+100}_{-80}$ Myr.}
    \label{fig:stellar}
\end{figure}

From the $a/R_s$ measured from the white light curve as described in Section~\ref{sec:methods}, we calculate the stellar density to be $\rho_s = 0.589 \pm 0.0038 $~g~cm$^{-3}$. We plot the results of these analyses in Figure~\ref{fig:stellar}. Our final age inference of $140_{-80}^{+100}$ Myr indicates the WASP-178 is relatively young compared to the Solar System and only a fraction into  the host star's expected lifetime of about 1.7 Gyr given its estimated mass \citep{harwit:1988}. We discuss these system parameters in the context of WASP-178b's formation in Section~\ref{sec:formation}.

As argued by \cite{reggiani:2022}, the atmospheric elemental abundances of giant planets can only be interpreted relative to the elemental abundances of their parent protoplanetary disks as often observable in the photospheres of their host stars.  This approach is not applicable for WASP-178 b, as the star WASP-178 is known to be a chemically peculiar A star of subclass Am \citep{hellier:2019} for which the abundances in its photosphere are not representative of its bulk abundances or the primordial abundances of the protoplanetary disk it once possessed. For chemically peculiar planet host stars like beta Pic that are members of a moving group, solar-type stars in the same moving group can be used as protoplanetary disk abundance proxies \citep{reggiani:2024}.  WASP-178 is not known to be a member a moving group, so some other approach is needed to estimate the elemental abundances of WASP-178 b’s parent protoplanetary disk.

The accurate and precise age ($\tau = 140^{+100}_{-80}$ Myr) we inferred for WASP-178 based on our measurement of its mean density allows us to use solar neighborhood open clusters of the same age to infer the expected elemental abundances of WASP-178.  To that end, we consider all open clusters with individual star membership data from \citep{gaia:2018b} and select those clusters within 500 pc of the Sun and ages consistent with age of WASP-178.  We then focus on the subset of those clusters with 15 or more published elemental abundance inferences for individual stars from either the fourth phase of the Sloan Digital Sky Survey (SDSS-IV) or the Gaia-ESO Survey (GES) and use them to infer the expected abundance of WASP-178 \citep{apogee:2022, ges:2022a, ges:2022b}.

The Pleiades and $\alpha$ Persei are the only two open clusters with SDSS-IV data and ages consistent with WASP-178.  We find that the average elemental abundances of those two clusters are consistent, supporting the fidelity of our approach. Blanco 1 and NGC 6405 are the only two open clusters with GES data and ages consistent with WASP-178.  We find that the average elemental abundances of those two clusters are consistent, further supporting the fidelity of our approach.  Moreover, these GES-based abundances are also consistent with the SDSS-IV based abundances of the Pleiades and alpha Persei thereby confirming the suitability of young open clusters as proxies for the unknown abundances of the WASP-178’s former protoplanetary disk. 

Overall, combining both the SDSS-IV and GES data, we find the bulk abundances of WASP-178, and therefore the protoplanetary disk from which WASP-178 b was formed, which are shown in Table \ref{tab:orb_params}. All elemental ratios measured are less than 0.15 dex within solar composition. Throughout the rest of the paper, we compare abundances in the planet's atmosphere relative to stellar and solar primordial abundances interchangeably based on these results.

\section{Data Reduction and Analysis}\label{sec:methods}

\subsection{JWST/NIRSpec/G395H Observations}

We observed one transit of WASP-178b with JWST/NIRSpec/G395H as part of Cycle 1 program JWST-GO-2055 (PI: Lothringer) on March 7, 2023. The visit consisted of a 7-hour time-series, centered on the 3.47 hour-long transit of WASP-178b. We used the SUB2048 subarray to read out both detectors, NRS1 and NRS2, obtaining the full wavelength coverage of 2.8--5.1~$\mu$m. We used the NRSRAPID readout mode with 605 total integrations of 45 groups for a total of 27,225 frames
, reaching a maximum of 80\% saturation in the brightest pixel.

 For target acquisition, due to the high saturation magnitude (J=11.2) of the fastest NIRSpec Wide Aperture Target Acquisition sequence, we did not directly acquire on WASP-178 (J=9.775, \citealt{skrutskie:2006}) as this would fully saturate $\geq 2$ pixels, which could prevent the target acquisition algorithm from working. We therefore first acquired on the nearby (17.867 arcseconds) J=16.3 star 15090619-4242069 \citep{gaia:2018} before offsetting to the location of WASP-178.

\subsubsection{\texttt{FIREFLy} Reduction}

\begin{deluxetable*}{c|c|c|c|c|c}
\tablecaption{WASP-178 System Parameters}\label{tab:orb_params}
\tablehead{
\colhead{Parameter} & \colhead{NRS1} & \colhead{NRS2} & \colhead{Weighted Average} & \colhead{Joint Fit} & \colhead{Reference}
}
\startdata
\multicolumn{6}{c}{\textbf{Planetary Parameters\tablenotemark{a}}} \\
\hline
$T_0$ & $0.870808 \pm 0.000026$ & $0.870851 \pm 0.000035$ & $0.870826 \pm 0.00002$ & $0.87080 \pm 0.00003$ & This work \\
$a/R_s$ & $7.037 \pm 0.018$ & $7.037 \pm 0.025$ & $7.037 \pm 0.015$ & $7.037 \pm 0.015$ & This work \\
$b$ & $0.5418 \pm 0.0033$ & $0.5398 \pm 0.0048$ & $0.5410 \pm 0.0027$ & $0.5420 \pm 0.0030$ & This work \\
$(R_p/R_s)^2$ (\%) & $1.1512 \pm 0.0012$ & $1.1537 \pm 0.0016$ & $1.1522 \pm 0.001$ & - & This work \\
$q_1$ (fixed) & 0.0198 & 0.0131 & - & - & This work \\
$q_2$ (fixed) & 0.1355 & 0.1152 & - & - & This work \\
\hline
$R_p$ ($R_J$) & \multicolumn{4}{c}{$1.77\pm 0.01$} & This work\\
$M_p$ ($M_J$) & \multicolumn{4}{c}{$1.66\pm 0.12$} & \cite{hellier:2019} \\
$\log{g}$ (cm/s$^2$) & \multicolumn{4}{c}{$3.12\pm 0.03$} & This work\\
Spin-Orbit $\lambda$ ($^\circ$) & \multicolumn{4}{c}{$105^{+3.6}_{-4.2}$} & \cite{damasceno:2024}\\
\hline
\multicolumn{6}{c}{\textbf{Stellar Parameters}} \\
\hline
$T_{\mathrm{eff}}$ (K) & \multicolumn{4}{c}{$9200^{+200}_{-170}$} & This work \\
$\log(g)$ (cgs) & \multicolumn{4}{c}{$4.28 \pm 0.01$} & This work \\
$\mathrm{[Fe/H]}$ (dex) & \multicolumn{4}{c}{$0.03^{+0.10}_{-0.12}$} & This work \\
$M_s$ ($M_\odot$) & \multicolumn{4}{c}{$2.02^{+0.04}_{-0.05}$} & This work \\
$R_s$ ($R_\odot$) & \multicolumn{4}{c}{$1.70 \pm 0.01$} & This work \\
$\rho_s$ (g cm$^{-3}$) & \multicolumn{4}{c}{$0.589 \pm 0.0038$} & This work \\
Age (Myr) & \multicolumn{4}{c}{$140^{+100}_{-80}$} & This work \\
$A_V$ (mag) & \multicolumn{4}{c}{$0.18^{+0.04}_{-0.05}$} & This work \\
$d$ (pc) & \multicolumn{4}{c}{$408 \pm 3$} & This work \\
$v\sin{i}$ (km s$^{-1}$) & \multicolumn{4}{c}{$8.2 \pm 0.6$} & 
\cite{hellier:2019} \\
$V_{\mathrm{turb}}$ (km s$^{-1}$) & \multicolumn{4}{c}{$2.9 \pm 0.2$} & \cite{hellier:2019} \\
Variability (mmag) & \multicolumn{4}{c}{$<1.5$} & \cite{hellier:2019} \\
$\mathrm{[Fe/H]_{phot}}$ (dex)  & \multicolumn{4}{c}{$+0.21 \pm 0.16$} & \cite{hellier:2019} \\
$\mathrm{[Ca/H]_{phot}}$ (dex) & \multicolumn{4}{c}{$-0.06 \pm 0.14$} & \cite{hellier:2019} \\
$\mathrm{[Sc/H]_{phot}}$ (dex)  & \multicolumn{4}{c}{$-0.35 \pm 0.08$} & \cite{hellier:2019} \\
$\mathrm{[Cr/H]_{phot}}$ (dex)  & \multicolumn{4}{c}{$+0.43 \pm 0.10$} & \cite{hellier:2019} \\
$\mathrm{[Ni/H]_{phot}}$ (dex)  & \multicolumn{4}{c}{$+0.32 \pm 0.12$} & \cite{hellier:2019} \\
\hline
\multicolumn{6}{c}{\textbf{Open Cluster Proxy Abundance Analysis}} \\
\hline
$\mathrm{[Fe/H]}$ & \multicolumn{4}{c}{$-0.02 \pm 0.06$} & This work \\
$\mathrm{[Al/Fe]}$ & \multicolumn{4}{c}{$-0.04 \pm 0.05$} & This work \\
$\mathrm{[C/Fe]}$ & \multicolumn{4}{c}{$-0.10 \pm 0.09$} & This work \\
$\mathrm{[Ca/Fe]}$ & \multicolumn{4}{c}{$-0.01 \pm 0.07$} & This work \\
$\mathrm{[Cr/Fe]}$ & \multicolumn{4}{c}{$-0.06 \pm 0.07$} & This work \\
$\mathrm{[K/Fe]}$ & \multicolumn{4}{c}{$-0.04 \pm 0.05$} & This work \\
$\mathrm{[Mg/Fe]}$ & \multicolumn{4}{c}{$-0.07 \pm 0.03$} & This work \\
$\mathrm{[Mn/Fe]}$ & \multicolumn{4}{c}{$+0.03 \pm 0.08$} & This work \\
$\mathrm{[N/Fe]}$ & \multicolumn{4}{c}{$+0.13 \pm 0.07$} & This work \\
$\mathrm{[Ni/Fe]}$ & \multicolumn{4}{c}{$-0.05 \pm 0.04$} & This work \\
$\mathrm{[O/Fe]}$ & \multicolumn{4}{c}{$+0.01 \pm 0.04$} & This work \\
$\mathrm{[Sc/Fe]}$ & \multicolumn{4}{c}{$-0.06 \pm 0.07$} & This work \\
$\mathrm{[Si/Fe]}$ & \multicolumn{4}{c}{$+0.01 \pm 0.02$} & This work \\
$\mathrm{[Ti/Fe]}$ & \multicolumn{4}{c}{$-0.06 \pm 0.04$} & This work \\
\enddata
\tablenotetext{a}{Orbital Parameters of WASP-178b are taken from the JWST-only white light curve fits with \texttt{FIREFLy}.}
\end{deluxetable*}

We reduced the data using the Fast InfraRed Exoplanet Fitting Lyghtcurve (\texttt{FIREFLy})
\citep{rustamkulov:2022, sing:2024} reduction suite. The reduction started with the uncalibrated (\texttt{uncal.fits})
images and performs a customized reduction using the STScI \texttt{jwst} pipeline. During stage 1, we include 1/$f$ destriping at the group level before the ramp is fit. 
We used the default superbias step, but skipped the dark-current stage of the STScI pipeline. We then use the custom-run pipeline 2D images after the \texttt{assign\_wcs} step, and performed customized cleaning of bad pixels, cosmic rays and hot pixels identifying them both spatially and in time. These corrections affect about 0.002\% of the pixels spatially and 0.02\% in time for the two detectors. A further 1/$f$ destriping is also performed at the integration level. We optimized the width of our flux extraction aperture to minimize the standard deviation of the white light curve, finding an aperture full-width of 5.3 pixels which was used to extract the 1D spectra with intrapixel flux extraction centered on the spectral trace.

 We then form a time series from the extracted 1D spectra, trimming off the first 575 and last 10 pixels of NRS1, as well as the first 8 and last 18 pixels of NRS2. The beginning of the time series appears well behaved so we do not trim any frames at the beginning of the time series. We then fit this white light curve of each detector (NRS1 and NRS2) independently, varying the transit center time (T$_0$), the orbital distance in units of stellar radii (a/$R_s$), the impact parameter (b), transit depth (R$_p$/R$_s$), and the flux baseline. Stellar limb darkening was fixed to quadratic coefficients fit to the custom WASP-178 PHOENIX stellar model used in \citepalias{lothringer:2022}. 

 In our fits to the white light curve, we tried a variety of systematics models for detrending, including the x and y position of the spectrum on the dector (as well as their covariate xy), an integration level scaling of the superbias, and trends with time. For NRS1, a linear trend in time was the only statistically significantly decorrelate (i.e., the only vector whose coefficient was not found to be consistent with zero). For NRS2, no detrending vectors were justified and effectively the raw extracted light curve was directly fit. These same systematics models (or lack thereof in the case of NRS2) were used for the spectroscopic light curves as well. Table~$\ref{tab:orb_params}$ lists the orbital parameters fit from the white light curves.\added{ We present orbital parameters as both the weighted average from the NRS1 and NRS2 fits, as well as a joint fit to both datasets simultaneously. We use the former in the spectroscopic fits described below.}

\begin{figure*}[t]
\gridline{\fig{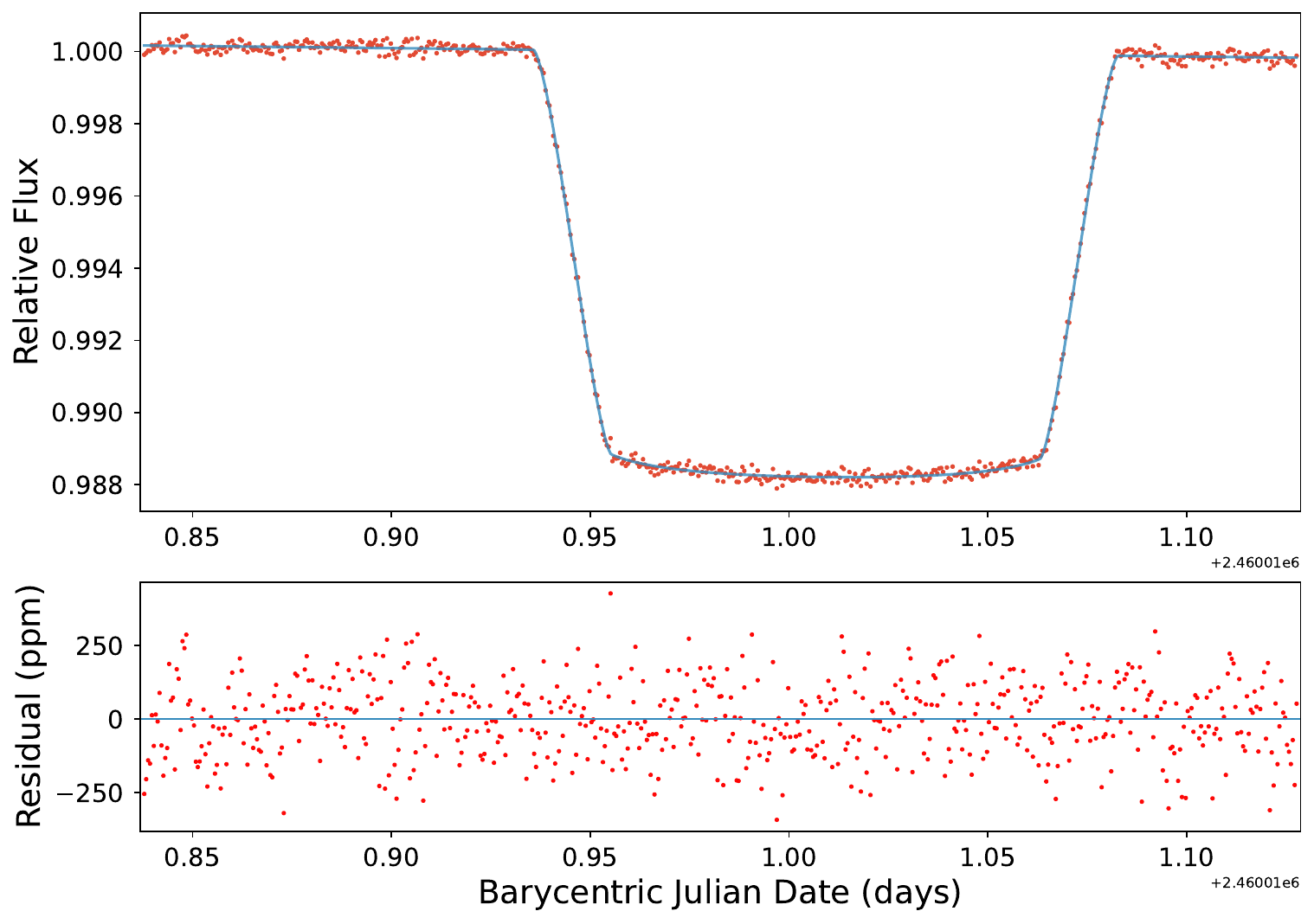}{0.49\textwidth}{(a) NRS1}
\fig{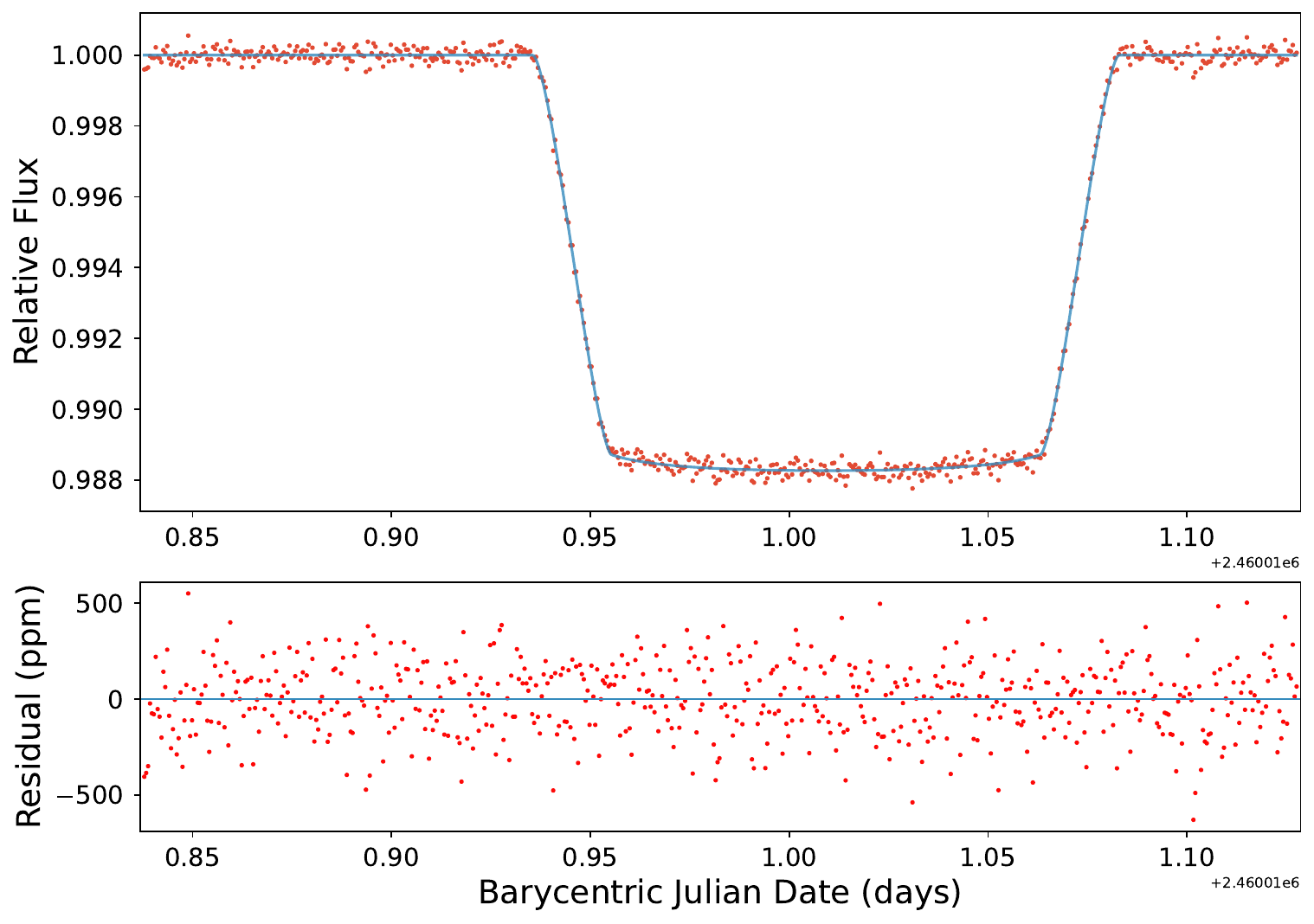}{0.49\textwidth}{(b) NRS2}}
\caption{White light curve fits to the \texttt{FIREFLy}-reduced JWST/NIRSpec/G395H data for NRS1 (left) and NRS2 (right). The top panel shows the fit to the raw data, normalized to the out-of-transit baseline. The bottom panel shows the resulting residuals to the fit. While the NRS1 light curve is fit with a linear slope in time, the NRS2 light curve is not. Significant red noise was not detected in the Allan deviation.}
\label{fig:wlc}
\end{figure*}

 We fit the light curves using the \texttt{emcee} MCMC sampler \citep{foreman-mackey:2012}. 
 For NRS1, our fit results in a scatter of about 125 ppm per point with a calculated photon noise of 101 ppm, meaning that our fit reaches 123\% of photon noise. For NRS2, the fit is a little worse with 185 ppm observed scatter with 132 ppm photon noise for 139\% photon noise. Figure~\ref{fig:wlc} shows the resulting fits to the data and their residuals. We find no sources of significant red noise when calculating the Allan deviation\added{ and therefore to did not pursue more complex systematics modelling of the lightcurves.}.

\begin{figure*}
    \centering
    \includegraphics[width=1.0\linewidth]{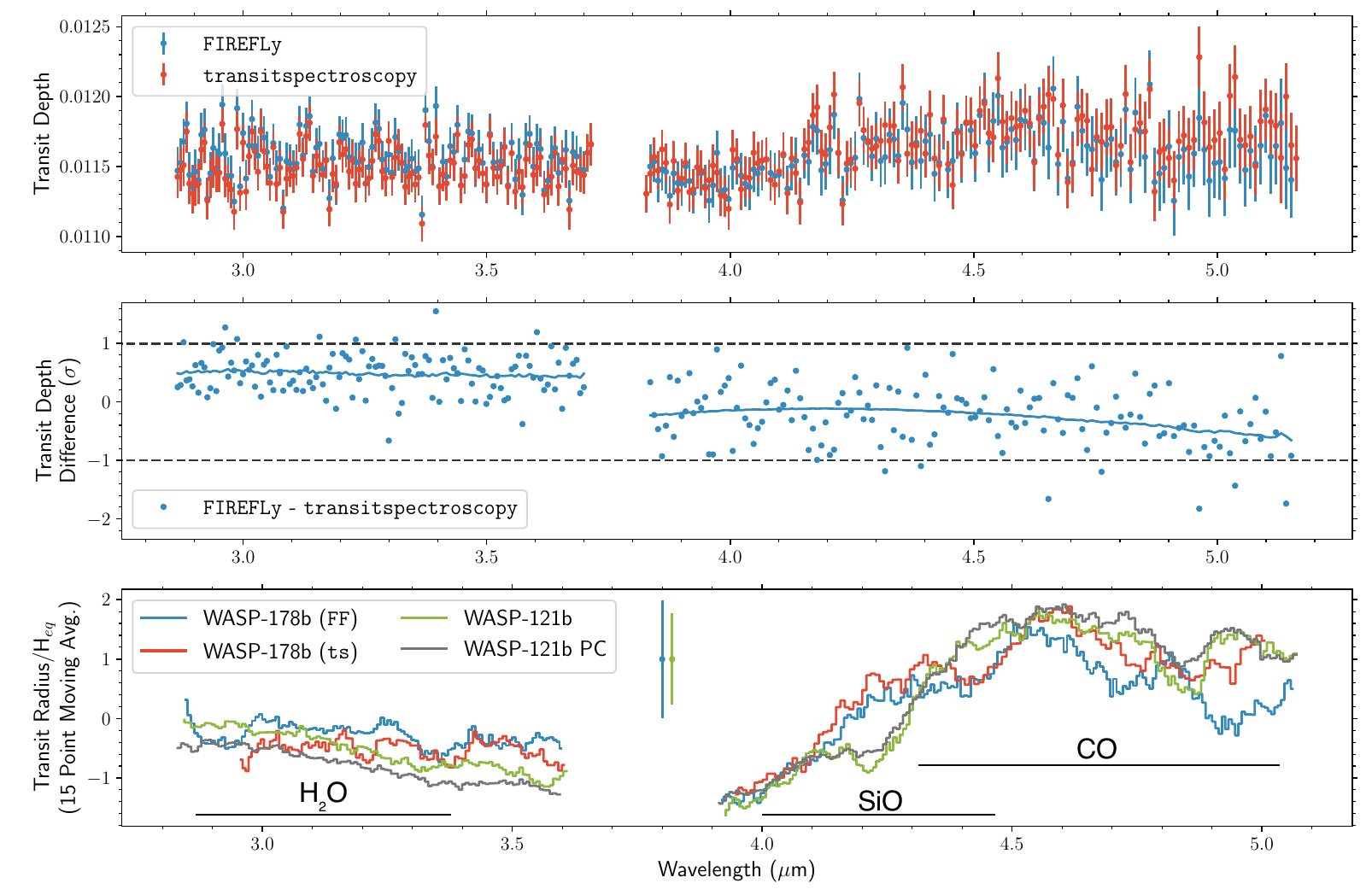}
    \caption{Top: Transmission spectrum of WASP-178b from the \texttt{FIREFLy} and \texttt{transitspectroscopy} reductions. Middle: Transit depth difference between the \texttt{FIREFLy} and \texttt{transitspectroscopy} reductions, in units of the transit depth uncertainty. Solid lines show a second order polynomial fit to NRS1 and NRS2. Bottom: The \texttt{FIREFLy} reduction compared to two versions of the transit spectrum of ultra-hot Jupiter WASP-121b from \cite{gapp:2025}, after normalizing to each planet's equilibrium scale height and taking a 15-point moving average. The points with errorbars indicate the average measurement uncertainty.}
    \label{fig:spec_compare}
\end{figure*}

 After fitting the white light curve, we split the data into spectroscopic time series with 10 pixel-wide bins for an effective resolution of approximately R$\sim$500. The two-dimensional spectroscopic light curves are shown in Appendix~\ref{appendix:2DLC} Figures~\ref{fig:NRS1_2D_LC} and \ref{fig:NRS2_2D_LC}. We used the same systematics model as in the white light curve fits and similarly fixed the quadratic limb-darkening coefficients to the custom WASP-178 PHOENIX stellar model \citep{hauschildt:1999} used in \citepalias{lothringer:2022}. \replaced{The light curve}{Due to their relatively Gaussian posteriors, the spectroscopic light curves} were then fit using weighted least-squares minimization with \texttt{lmfit} \citep{newville:2016} using orbital parameters fixed to the weighted average of the NRS1 and NRS2 values. The resulting spectroscopic transit depths are shown in Figure~\ref{fig:spec_compare}. We compared the transmission spectrum to an identical analysis with freely fit quadratic limb-darkening coefficients and found the spectra to be very similar (see Appendix~\ref{appendix:LC} Figures~\ref{fig:NRS1_LC_fit} and \ref{fig:NRS2_LC_fit}). For atmospheric interpretation, described below, we chose the fixed limb-darkening reduction because it had somewhat smaller error bars and the results were consistent with the fitted limb-darkening analysis.

\subsubsection{\texttt{transitspectroscopy} Reduction}
We also independently reduced the WASP-178b JWST observations with \texttt{transitspectroscopy} \citep{espinoza:2022}.
Stage 1, a wrapper for the \texttt{jwst} calibration pipeline (v1.14.0), applied default calibration steps, replacing the jump detection step with a custom function that calculates differences between consecutive groups for each pixel. Stage 2 converted raw integration images into time-series light curves for both detectors. The spectral traces were determined using cross-correlation with a Gaussian profile and smoothed using a spline function. The trace spanned 550–2041 x-pixels for NRS1 and 5–2042 x-pixels for NRS2. The background subtraction is performed by masking the spectral trace in the median rate frame, calculating the median of out-of-trace pixels, and subtracting this background from each integration frame. For 1/f noise removal, the median frame was scaled to each individual integration and then subtracted --- from this residual frame, the 1/f noise in a given column is then  estimated as the median of masked column pixels outside a 3-pixel inner radius and beyond an 8-pixel outer radius. This noise was then subtracted from each integration. The stellar flux was extracted using a simple box method with a 3-pixel radius. We generated pixel light curves for each column and white light curves by summing flux across all pixels in each detector.

Light curves were fit using \texttt{juliet} \citep{espinoza:2019}, which incorporates the \texttt{batman} package \citep{kreidberg:2015} for modeling transits. Orbital parameters—period ($P$), mid-transit time ($t_0$), scaled semi-major axis ($a/R_{\star}$), and impact parameter (b)—were fixed to refined values from \citet{hellier:2019}. The planet-to-star radius ratio ($R_{P}/R_{\star}$) was fitted with uniform priors (0–0.2). Limb-darkening coefficients (\(u_1, u_2\)) for the quadratic law were sampled uniformly between -3 and 3. Additional parameters included a mean out-of-transit flux factor ($m_{\text{flux}}$; normal distribution, mean 0, standard deviation 0.1) and a white noise jitter term ($\sigma_w$; log-uniform, 10–10,000 ppm). Systematic trends were modeled with a linear function in time. The transit depths and their uncertainties were derived as the median and variance of the posterior distribution of \replaced{$R_{P}/R_{\star}^2$}{$(R_p/R_s)^2$} for each wavelength bin.

\subsection{HST/WFC3 Observations}

To extend the existing HST/WFC3/G280 (0.2--0.8~$\mu$m) transmission spectrum of WASP-178b, we observed one transit of WASP-178b with HST/WFC3/G102 (0.8--1.15 $\mu$m) on March 4th, 2021 and one transit with G141 (1.125--1.65 $\mu$m) on February 19th, 2021 as part of program HST-GO-16450 (PI: Lothringer). These observations use an HgCdTe detector and utilize an ``up-the-ramp" sampling technique, analogous to JWST. Both WFC3-IR observations utilized the spatial-scanning mode \citep[e.g.,][]{deming:2013}, in which forward and reverse scans are created as the telescope nods slowly in the cross-dispersion direction over the course of the observation. This allows for increased S/N by increasing the duty cycle of the observations and maximizing the amount of time collecting photons. The observations took place over the course of five orbits, with eight groups per integration. The G102 observation had 70 total integrations and the G141 observation had 78 total integrations. 

\subsubsection{HST/WFC3 IR Reduction}

To reduce these observations, we used the pipeline \texttt{WRECS} \citep{stevenson:2014c}, which is the precursor to the \texttt{Eureka!} pipeline \citep{bell:2022}. We began with the \texttt{ima} files and utilize ``difference images" in order to calculate the up-the-ramp slope. This was followed by background subtraction, after which we applied a rough pixel-drift correction. Next, we performed a $10\sigma$ outlier rejection, and finally applied a sub-pixel drift correction.

We extracted the stellar spectrum with an aperture half-width of 25 pixels using the optimal extraction algorithm implemented by \cite{horne:1986b}. To fit the white light and spectroscopic light curves, we assumed a circular orbit and fixed $T_0$, $a/R_s$, and $b$ to the fitted values from the JWST white-light curve fits (Table \ref{tab:orb_params}). We also fixed $P=3.3448$~days \citep{hellier:2019}. We used the quadratic limb darkening law and fixed the limb darkening coefficients using the PHOENIX stellar model from \citepalias{lothringer:2022} to keep the HST reduction consistent with the JWST reduction.

The transit itself was modeled using the analytic approach from \cite{mandel:2002}. The systematics were handled by fitting a linear slope alongside an exponential ramp, as is standard for HST/WFC3 IR observations (e.g., \citealt{berta:2012, kreidberg:2014, stevenson:2014c}). If the ramp is not fit and we only fit for the linear slope (as in \citealt{deming:2013}), we are left with correlated noise in the residuals. We fit the forward and reverse scans separately and excluded the first orbit in the white light curve fit due to the sharper ramp seen in the first orbit (e.g., see \citealt{berta:2012, kreidberg:2014}). For both the white light and spectroscopic light curves, we used Differential-Evolution Markov Chain Monte Carlo (DEMC) \citep{terbraak:2006,terbraak:2008} to fit the light curves.   

We use the best-fit white light model to apply a common-mode correction in the spectroscopic fits. This entails dividing each spectroscopic light curve by the transit-removed white light curve, which removes non-wavelength-dependent effects. In this way, we remove the exponential ramp and do not re-fit for it at the spectroscopic stage. The only systematic we fit for is the linear term, which does have some wavelength-dependence. For the spectroscopic fits, we used all five orbits, though we excluded the first six points from the first orbit, as these still show a strong ramp and greatly increase the red noise in the residuals. Additionally, we combine the forward and reverse scans together at this point. Other than that, we treat the spectroscopic fits analogously to the white light curve fits (fixing the same orbital parameters and using quadratic limb darkening from the same custom PHOENIX model of the star used for the G280 limb-darkening coefficients). Once the light curves are fit, we rescale the error bars if there is red noise in the residuals (calculated as the median of the excess noise across all bin sizes in time). The final WFC3/G102 and G141 transmission spectra are seen in Figure \ref{fig:spec_fits}.

We combined the HST WFC3/IR observations with the HST/WFC3/G280 transit spectrum of WASP-178b from program HST-GO-16086 (PI: Lothringer), described in \citepalias{lothringer:2022}. 

\subsection{The UVOIR Transmission Spectrum of WASP-178b}

Figure~\ref{fig:spec_compare} shows the reduced JWST observations with \texttt{FIREFLy} and \texttt{transitspectroscopy}. \added{NRS1 shows a slope towards short wavelengths corresponding to H$_2$O absorption, while NRS2 shows a stronger absorption peak at 4.6~$\mu$m, corresponding to CO.} No other major spectral features are clearly evident by eye, though both reductions do have smaller, higher-frequency peaks and troughs. While the self-consistent 1D-RCE PHOENIX model shown in Figure~\ref{fig:spec_fits} exhibits a similarly flat H$_2$O feature, the magnitude of the CO absorption is significantly over-predicted.

The two reductions roughly agree, with a divergence at the longest wavelengths and a sub-$\sigma$ offset between the NRS1 and NRS2 detectors, a difference that has been seen in other data sets with these same pipelines \citep{gressier:2024}, likely arising due to differences in the 1/$f$ noise subtraction. The reductions were truly independent, however, with only the bin locations matched for Figure~\ref{fig:spec_compare}. All other reduction steps were carried out independently, from the orbital parameters used to the limb-darkening coefficients. Despite this, as we discuss in Section~\ref{sec:results:full_rets:chemeq}, full-spectrum retrievals with each reduction give consistent interpretations, indicating that the atmospheric interpretations discussed here are relatively robust to the treatment of the data.

We also compare the spectrum of WASP-178b to that of WASP-121b in Figure~\ref{fig:spec_compare} \citep{gapp:2025}. Because the WASP-121b observations were taken as part of a phase curve, two reductions are shown, one fitting the out-of-transit baseline as flat and another using phase curve information to fit the out-of-transit baseline.
Compared to both reductions of the WASP-121b spectrum, WASP-178b's H$_2$O feature is significantly flatter. WASP-121b also exhibits shallow transit depths between 4.2 and 4.3~$\mu$m. WASP-178b shows no such trough at 4.2~$\mu$m and the spectrum instead smoothly increases towards the peak of the CO absorption around 4.6~$\mu$m. Even though the transit depths are very similar where SiO absorbs, there is no apparent IR spectral feature from SiO detected in WASP-178b. As we discuss in Section~\ref{sec:results:full_rets}, however, this lack of a spectral feature in the IR is still consistent with the large UV absorption being due to SiO.

Figure~\ref{fig:spec_fits} then shows the \texttt{FIREFLy} data reduction, used for the rest of our analysis, alongside the observations from HST/WFC3. The large NUV absorption dominates the entire transmission spectrum, stretching nearly 20 scale heights and dwarfing the H$_2$O and CO spectral features, which absorb over between 1 and 3 scale heights. The G102 and G141 spectral range is relatively featureless, with a downtrend in transit depth extending out to 1.6~$\mu$m indicative of strong H$^-$ opacity. This is another important difference between WASP-178b and WASP-121b, where the latter exhibited clear H$_2$O absorption at both 1.2 and 1.4~$\mu$m \citep{evans:2018}, while the former's spectrum is much more featureless.

\section{Atmosphere Modeling Methods} \label{sec:methods:retrieval}

In order to interpret our spectrum, we use a number of different approaches, including self-consistent 1D radiative convective modeling with PHOENIX, as well as atmospheric retrievals with both \pRT{} \citep{molliere:2019,nasedkin:2024} and \texttt{ATMO} \citep{drummond:2016,tremblin:2017,amundsen:2014}. Using \texttt{ATMO} allows us to directly compare to the results for WASP-121b \citep{gapp:2025}, which used an identical setup, while the use of \pRT{} enables us to retrieve a wide variety of short-wavelength opacity sources. We run retrievals with both chemical equilibrium and free chemistry and assume different temperature structure parametrizations. We describe each in turn below.

\subsection{PHOENIX Self-Consistent 1D Model}

To obtain our fiducial theoretical expectations, we calculated custom atmosphere models of WASP-178b with PHOENIX \citep[Version 19,][]{hauschildt:1999,barman:2001}. We use a similar setup to \citep{lothringer:2018b} with 64 vertical layers, spaced log-uniformly between an optical depth of $10^3$ and $10^{-10}$, corresponding to between approximately 70 and $10^{-10}$ bars, respectively. Our opacity includes all atoms and ions up to uranium from \cite{kurucz:1995}.
This version of PHOENIX uses updated molecular line lists compared to previous versions used to study exoplanets, particularly CO \citep{li:2015}, H$_2$O \citep{polyansky:2018}, SiO \citep{yurchenko:2022}, VO \citep{mckemmish:2016}, and TiO \citep{mckemmish:2019}. The model is irradiated with a 9400~K star with stellar parameters from \cite{hellier:2019}. We run the model iteratively, assuming full heat-redistribution, until it reaches within $\sim$1~K of radiative equilibrium in each layer.

\subsection{petitRADTRANS Retrieval}

We use petitRADTRANS \citep[Version 3,][]{molliere:2019,nasedkin:2024} to retrieve the atmospheric properties of WASP-178b from our observations. Our 1-dimensional setup uses 80 vertical layers spread evenly in log-space between 100 and $10^{-6}$ bars. We use pre-tabulated correlated-K tables to calculate opacities. Included molecular opacities were H$_2$O \citep{rothman:2010}, CO \citep{rothman:2010}, CO$_2$ \citep{ExoMolCO2}, SiO \citep{ExoMolSiO}, TiO \citep{mckemmish:2019}, VO \citep{mckemmish:2016}, and FeH \citep{wende:2010}. We also included Fe, Fe+, Mg, and Mg+ from \cite{kurucz:1995}. H$_2$-H$_2$ and H$_2$-He collision-induced absorption (CIA) is included, as well as H$^-$ \citep{gray:1992}, which is calculated in chemical equilibrium in all cases.

We performed retrievals both in chemical equilibrium, with O/H, C/H, Si/H, Fe/H, and Ti/H as free parameters (with the latter two fixed to solar for retrievals of only the JWST data), and with free chemistry, where the individual abundances are treated as vertically-uniform free parameters. We compared isothermal retrievals with retrievals that used the 3-parameter radiative-equilibrium parametrization from \cite{guillot:2010}, hereafter referred to as ``3TP". We tested the inclusion of a grey cloud layer in a chemical equilibrium retrieval as well, but its inclusion was not statistically justified. We ran the retrievals with the Multinest sampling algorithm \citep{fer08,fer09,fer19,buckner:2014}, with most retrievals using 1000 live points unless otherwise stated. Our priors are listed in Table~\ref{tab:priors}.

\begin{table}[h]
    \centering
    \begin{tabular}{lcc}
        \hline
        Parameter & Lower Bound & Upper Bound \\
        \hline
        Free Mass Fractions & $-10.0$ & 0.0 \\
        $\log_{10}\mathrm{[O/H]}$ & $-2.0$ & $-8.0$ \\
        $\log_{10}\mathrm{[C/H]}$ & $-2.0$ & $-8.0$ \\
        $\log_{10}\mathrm{[Si/H]}$ & $-3.0$ & $-9.0$ \\
        $\log_{10}\mathrm{[Fe/H]}$ & $-3.0$ & $-9.0$ \\
        $\log_{10}\mathrm{[Ti/H]}$ & $-5.0$ & $-11.0$ \\
        WFC3-IR offset & $-0.0010$ & 0.0010 \\
        WFC3-UVIS offset & $-0.0010$ & 0.0010 \\
        R$_\mathrm{p}$ (cm) & 8.73888$\times10^9$ & 1.71282$\times10^{10}$ \\
        $T_\mathrm{iso}$ (K) & 1000.0 & 4000.0 \\
        $T_\mathrm{eq}$ (K) & 1000.0 & 4000.0 \\
        $\log_{10}{(\gamma)}$ & $-2.0$ & 2.0 \\
        $\log_{10}{(\kappa_\mathrm{IR})}$ (cm$^2$/g) & $-3.0$ & 2.0 \\
        \hline
    \end{tabular}
    \caption{\pRT{} retrieval uniform priors for selected parameters.}
    \label{tab:priors}
\end{table}

We performed retrievals with the JWST data alone, as well as in combination with the three HST/WFC3 grism observations. When including the HST observations, we fit for two offsets between HST/WFC-UVIS/G280, HST/WFC-IR/G102+G141, and JWST/NIRSpec/G395H. We chose to fit for offsets between the non-overlapping datasets to marginalize over any uncertainty in the transit depth between instruments/detectors or between epochs of observations. To the latter point, while WASP-178 is not known to be a particularly active star, our observations are separated by a little more than two years, over which time there has been no known monitoring of the host star.

\subsection{\texttt{ATMO} Retrieval}

We also used the 1D-2D radiative-convective equilibrium model \texttt{ATMO}   \citep{amundsen:2014,drummond:2016,tremblin:2017} to retrieve the atmospheric properties of WASP-178b from our JWST observations. Additional details of the \texttt{ATMO} retrieval suite can be found in \cite{sing:2024}. \added{We did not run \texttt{ATMO} on the full HST+JWST dataset because of a lack of up-to-date atomic opacities at short wavelengths, relevant to fit the UVIS/G280 data. This would also require further developments to the chemistry implementation within \texttt{ATMO} to incorporate the atomic and ionized species. Therefore we used \texttt{ATMO} primarily as a check on agreement with \pRT{} for the JWST observations.}

For the NIRSpec WASP-178~b data, we fit the P-T profile with a parameterized \cite{guillot:2010} profile with two optical channels (``5TP") \cite{line:2012}, finding it preferred over a simple isotherm.
We assumed chemical equilibrium, including rainout. 
We fit the elemental abundances of  [C/C\textsubscript{$\odot$}], [O/O\textsubscript{$\odot$}], and [Si/Si\textsubscript{$\odot$}]
individually, with the rest of non-H/He elements varied together in a single metallicity term $Z$. These four elemental abundance parameters were used with the parameterized P-T profile to determine the chemical equilibrium abundances at each model evaluation. 
We included the spectrally active species of H$_2$-H$_2$,  H$_2$-He (CIA) opacities, 
H$_2$O \citep{barber:2006}, 
CO$_2$ \citep{tashkun:2011}, 
CO \citep{rothman:2010}, 
CH$_4$ \citep{yurchenko:2014}, 
NH$_3$ \citep{yurchenko:2011}, 
H$_2$S \citep{rothman:2010}, 
SO$_2$ \citep{underwood:2016},
SiO \citep{ExoMolSiO},
PH$_3$ \citep{Sousa-Silva:2015},
H- \cite{john:1988,bell:1987}
HCN \cite{barber:2014},
and
C$_2$H$_2$ \cite{rothman:2013}.
The fit consisted of 10 free parameters, 5 for the T-P profile, 4 for the abundances, and the planetary radius at a reference pressure of 1 mbar was also fit.

\begin{figure*}[th!]
    \centering
    \includegraphics[width=1.0\linewidth]{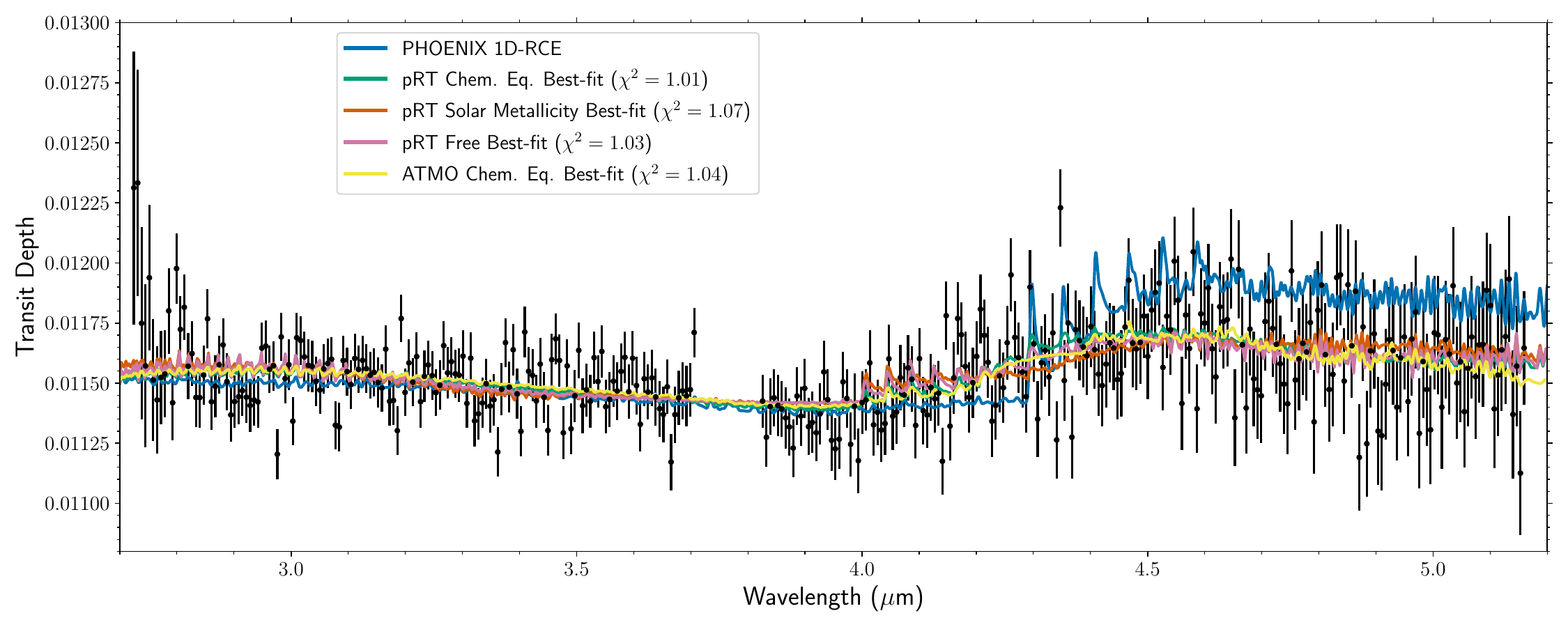}
    \includegraphics[width=1.0\linewidth]{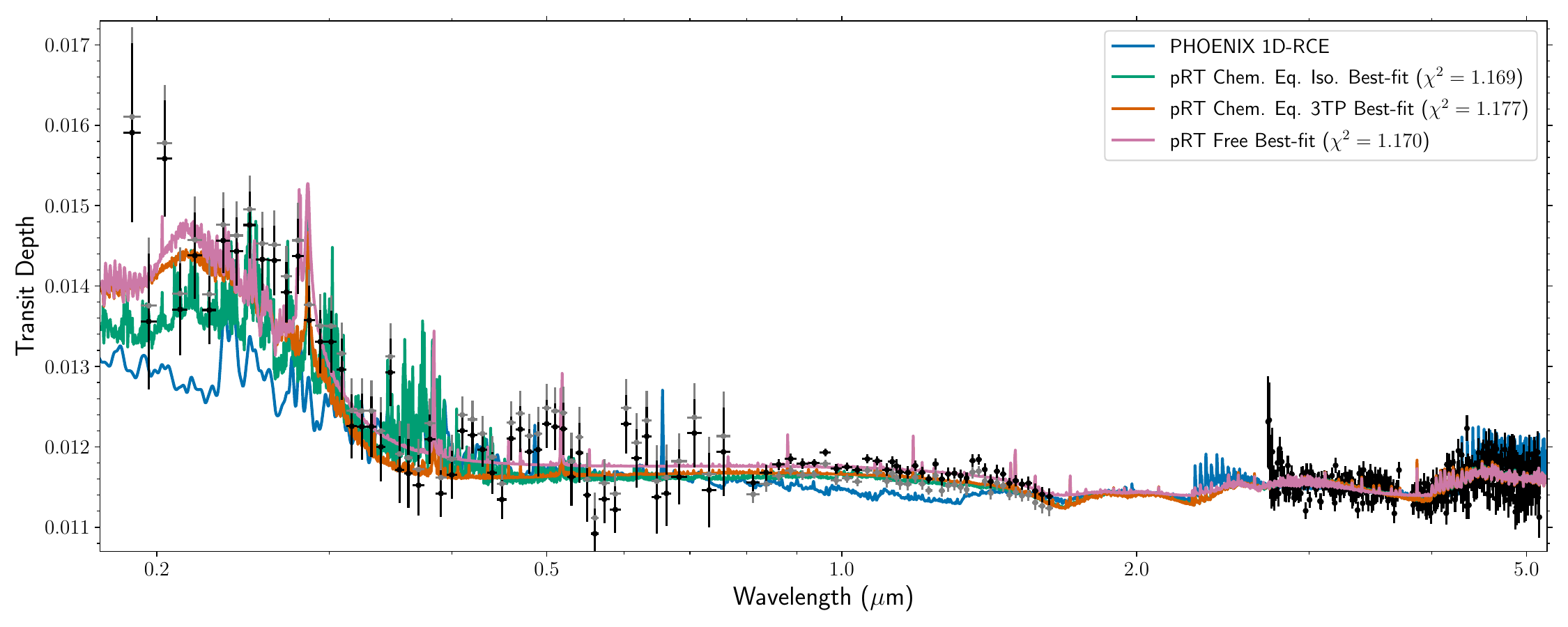}
    \caption{Top: Best-fit spectra from the \pRT{}\added{ (pRT)} and \texttt{ATMO} retrievals fitting only the JWST/NIRSpec/G395H spectrum. Bottom: Best-fit spectra from \pRT{} fitting the entire 0.2-5.1~$\mu$m HST+JWST spectrum. Black points show the spectrum with the offsets fitted in the free chemistry retrieval, while grey points show the spectrum with the offsets corresponding to the chemical equilibrium retrieval. While the self-consistent \added{PHOENIX radiative-convective equilibrium }model shows the correct features, it underpredicts the NUV absorption and overpredicts magnitude of the IR features. The flexibility of the retrievals allows them to fit the data well (i.e., $\chi^2/N < 1.2$).}
    \label{fig:spec_fits}
\end{figure*}

\section{Atmosphere Retrieval Results} \label{sec:results}

\subsection{JWST-Only Retrievals}\label{sec:results:jwstonly}

We first computed atmosphere retrievals using only the JWST/NIRSpec/G395H observations to investigate the precision with which we could measure H$_2$O and CO with IR data alone, as well as to search for additional absorption from species like SiO or CO$_2$.

\subsubsection{JWST-Only Chemical Equilibrium Retrievals}\label{sec:results:jwstonly:chemeq}

We performed atmospheric retrievals assuming chemical equilibrium using both \pRT{} and \texttt{ATMO}. The top panel of Figure~\ref{fig:spec_fits} shows the fits to the JWST data alone. Both the \pRT{} and \texttt{ATMO} chemical equilibrium retrievals fit the data well, reaching $\chi^2/N_{\mathrm{data}}$ of 1.01 and 1.04, respectively. 

\begin{figure*}[t]
    \centering
    \includegraphics[width=1.0\linewidth]{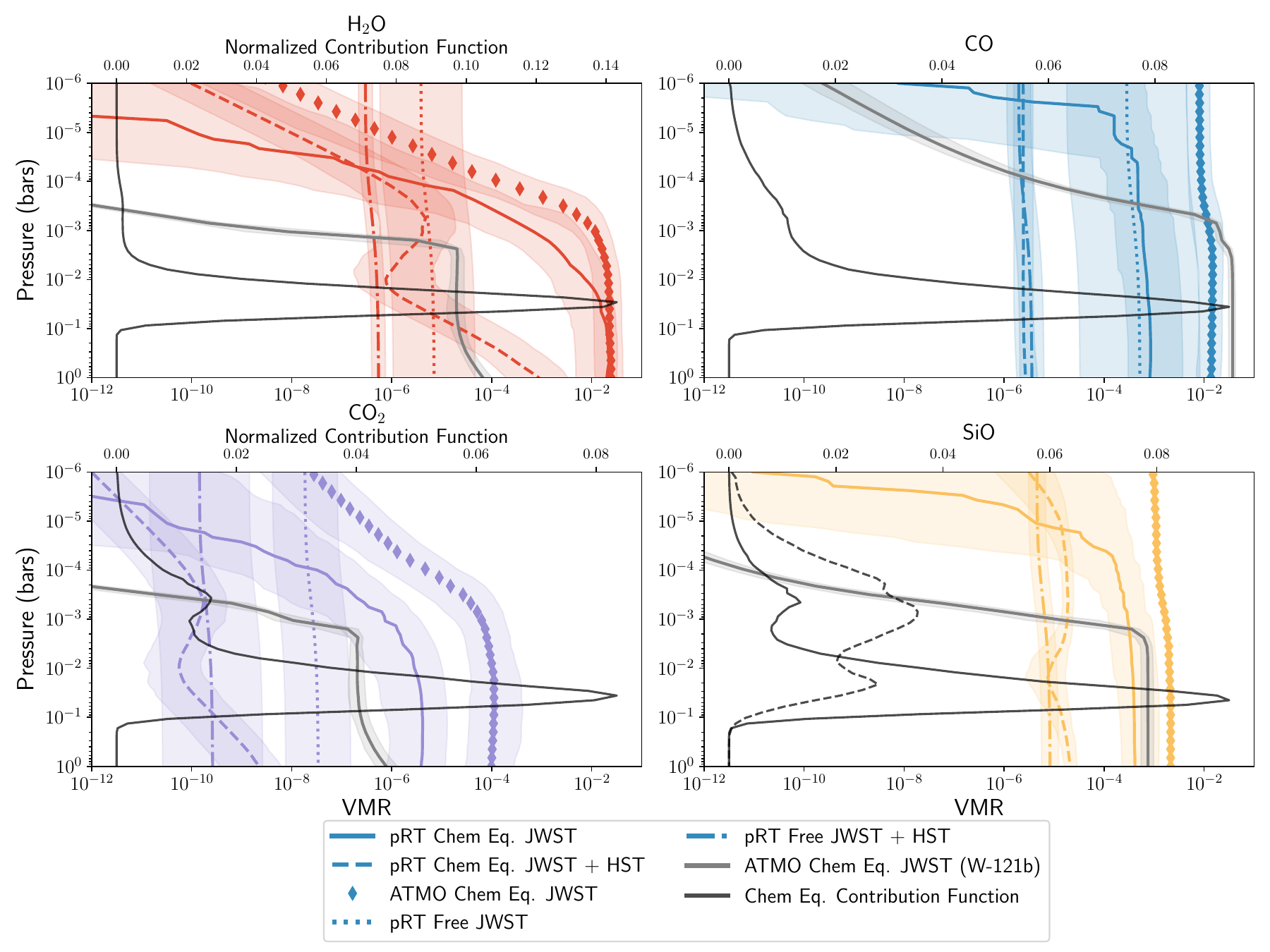}

    \caption{Abundance profiles for H$_2$O, CO, SiO, and CO$_2$ (clock-wise from upper-left) from the various retrieval scenarios. The abundance profile for WASP-121b from the \texttt{ATMO} chemical equilibrium retrieval is shown for comparison. Also shown are the normalized contribution functions (black) from the \texttt{pRT} chemical equilibrium retrieval, showing approximately where in the atmosphere each molecule is being measured. For SiO, the solid black contribution function shows where the 4.1~$\mu$m feature probes, while the dashed line shows where the NUV feature probes. The profile for the \texttt{ATMO} chemical equilibrium retrieval of SiO (dashed yellow) represents the 1-$\sigma$ upper-limit.}
    \label{fig:vmrs}
\end{figure*}

\begin{figure*}[t]
    \centering
    \includegraphics[width=0.95\linewidth]{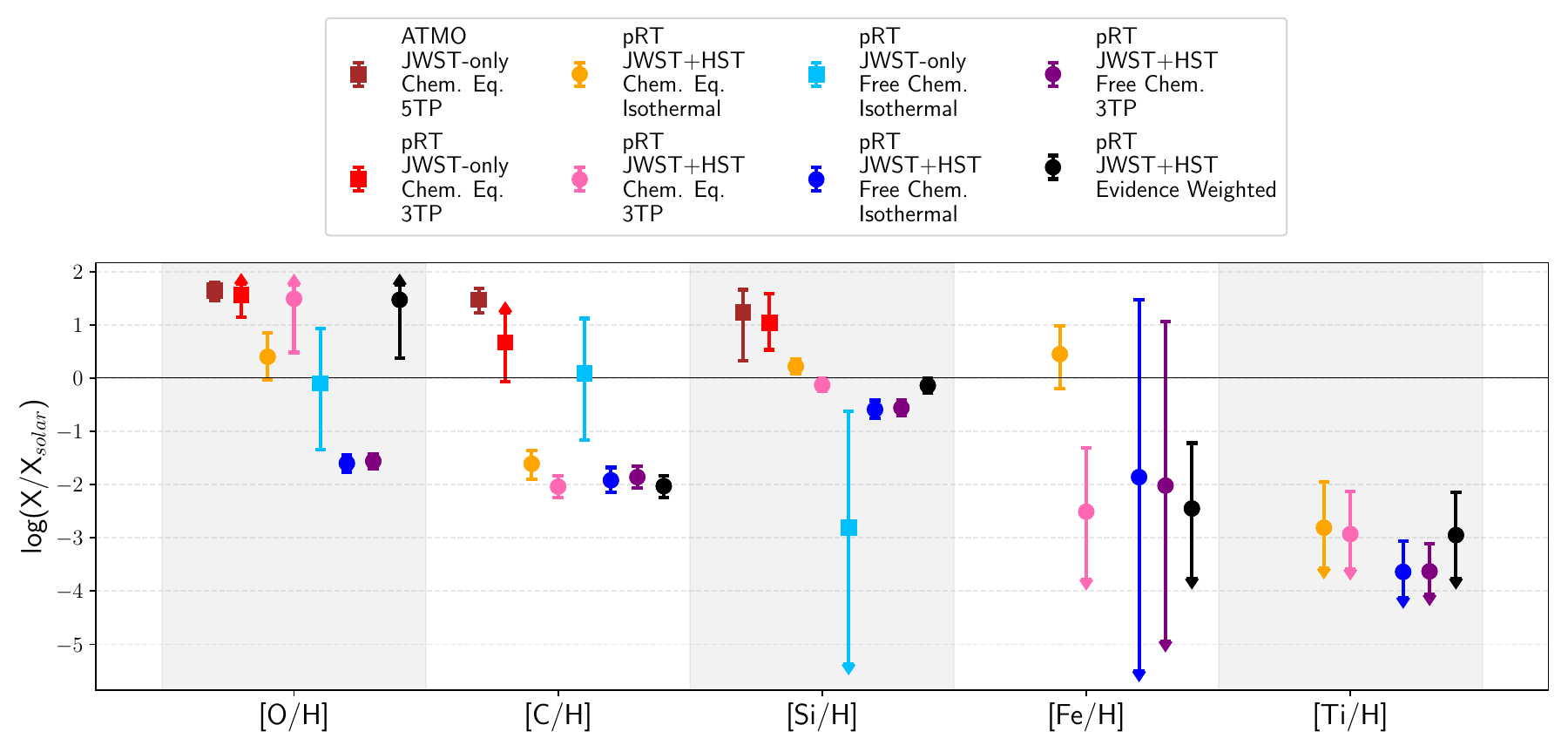}
    \includegraphics[width=0.712\linewidth]{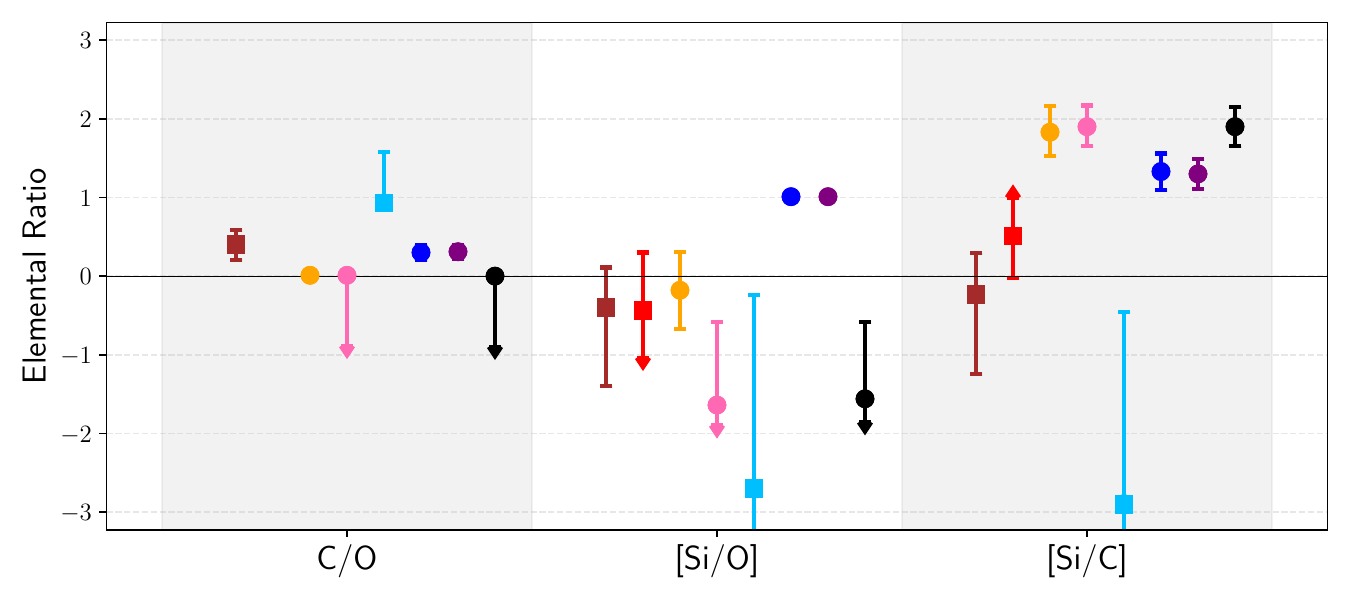}

    \caption{Graphical representation of the elemental ratios inferred from the retrievals listed in Table~\ref{table:retrieval} in the style of \cite{gandhi:2023}'s Figure~3. Redder colors represent retrievals assuming chemical equilibrium, while bluer colors represent free chemistry retrievals. Measurements with square markers retrieve on only the JWST data, while circular markers are measurements from retrievals of the full-spectrum dataset. In black we show our preferred constraints, the evidence-weighted measurements from all HST+JWST retrievals, dominated by the 3TP chemical equilibrium model in pink. \label{fig:abund_graphic}}
    \label{fig:enter-label}
\end{figure*}

Table~\ref{table:retrieval} lists the retrieved abundances from both \pRT{} and \texttt{ATMO} and Figure~\ref{fig:vmrs} shows the vertical profile of the retrieved abundances for H$_2$O, CO, CO$_2$, and SiO as implied by the retrievals. The elemental abundance ratios are also shown graphically in Figure~\ref{fig:abund_graphic}. The retrievals are both drawn to super-solar metallicities, with high [O/H] at $36^{+21}_{-22}$ and $44^{+26}_{-16}$ $\times$ solar, from \pRT{} and \texttt{ATMO} respectively. Each retrieval also measures a super-solar [Si/H] at $11^{+27}_{-8}$ and $17^{+26}_{-15}$ $\times$ solar for \pRT{} and \texttt{ATMO} respectively. We look at whether SiO is detected in the IR with the free retrievals in Section~\ref{sec:results:jwstonly:free}.

\begin{figure}[t]
    \centering
    \includegraphics[width=1.0\linewidth]{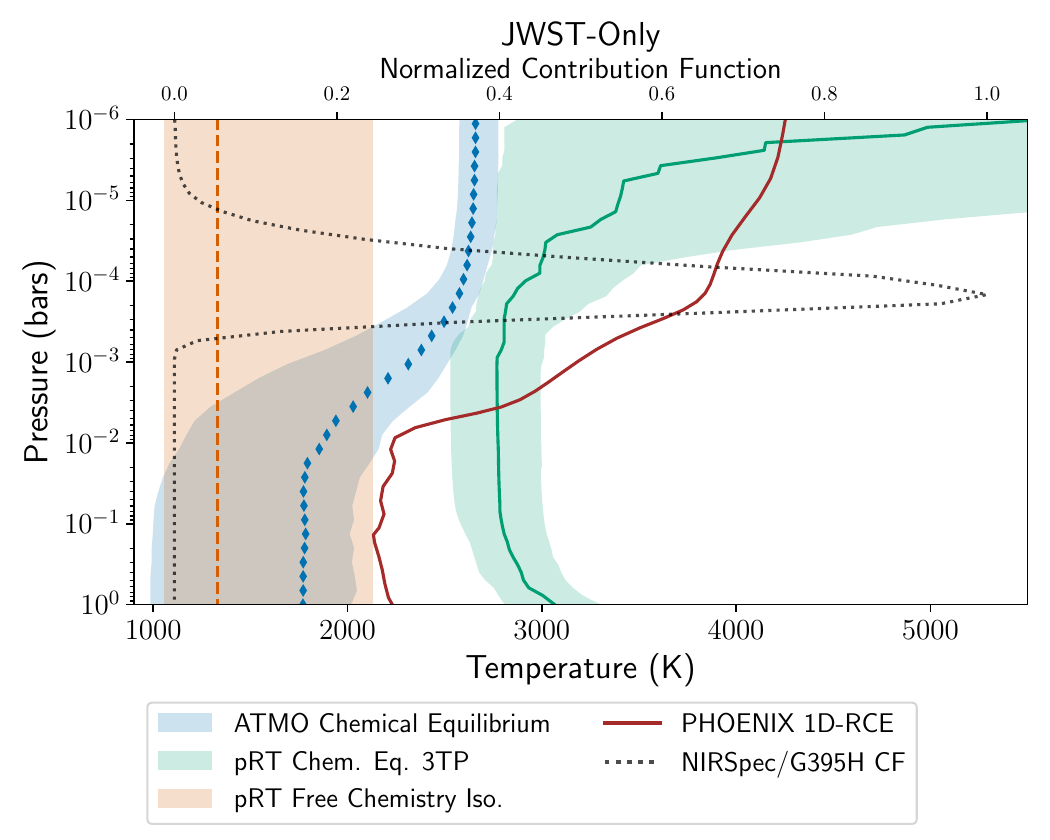}
    \includegraphics[width=1.0\linewidth]{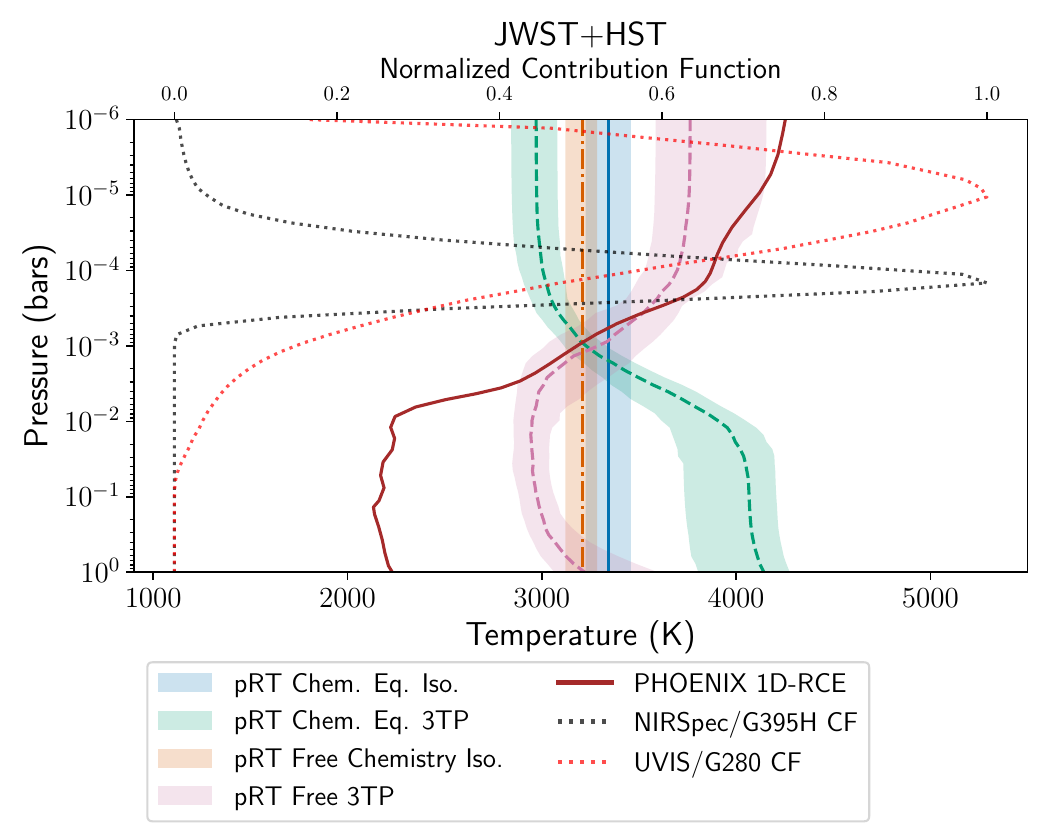}
    \caption{Temperature profiles from various retrieval scenarios and the PHOENIX 1D radiative-convective equilibrium model. Top: Retrievals using only the JWST NIRSpec observations. Bottom: Retrievals using both the HST and JWST observations.\added{ Dotted lines indicate relevant contribution function profiles.} While the shape of the retrieved temperature profile greatly varies, JWST+HST retrievals converge on a 1 mbar temperature around 3200~K. While this is much hotter than the planet's equilibrium temperature of $\sim$2450~K \citep{hellier:2019,rodriguezmartinez:2020}, this may indicate that the transmission spectrum probes a temperature inversion.}
    \label{fig:TP}
\end{figure}

While \texttt{ATMO} prefers a somewhat higher C/O than \pRT{} at $0.4\pm 0.19$ versus $0.09^{+0.25}_{-0.07}$, the 1$\sigma$ regions for both retrievals overlap at a slightly sub-solar C/O ratio. The difference in C/O ratio between the two retrievals may be attributed to the retrieved temperature structure in each retrieval, seen in Figure~\ref{fig:TP}, where \pRT{} prefers a hotter atmosphere compared to \texttt{ATMO}. Therefore, to make up for dissociated H$_2$O, \pRT{} must decrease the C/O ratio to preserve similar H$_2$O/CO ratios. 

\begin{figure*}
    \centering
    \includegraphics[width=0.48\linewidth]{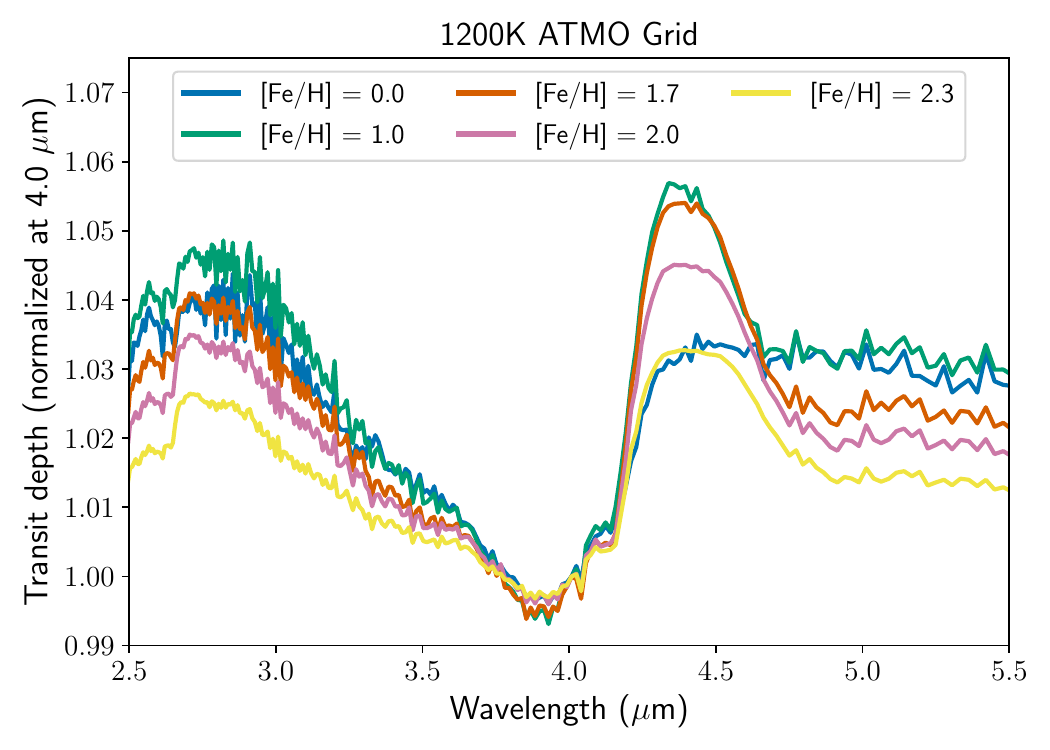}
    \includegraphics[width=0.48\linewidth]{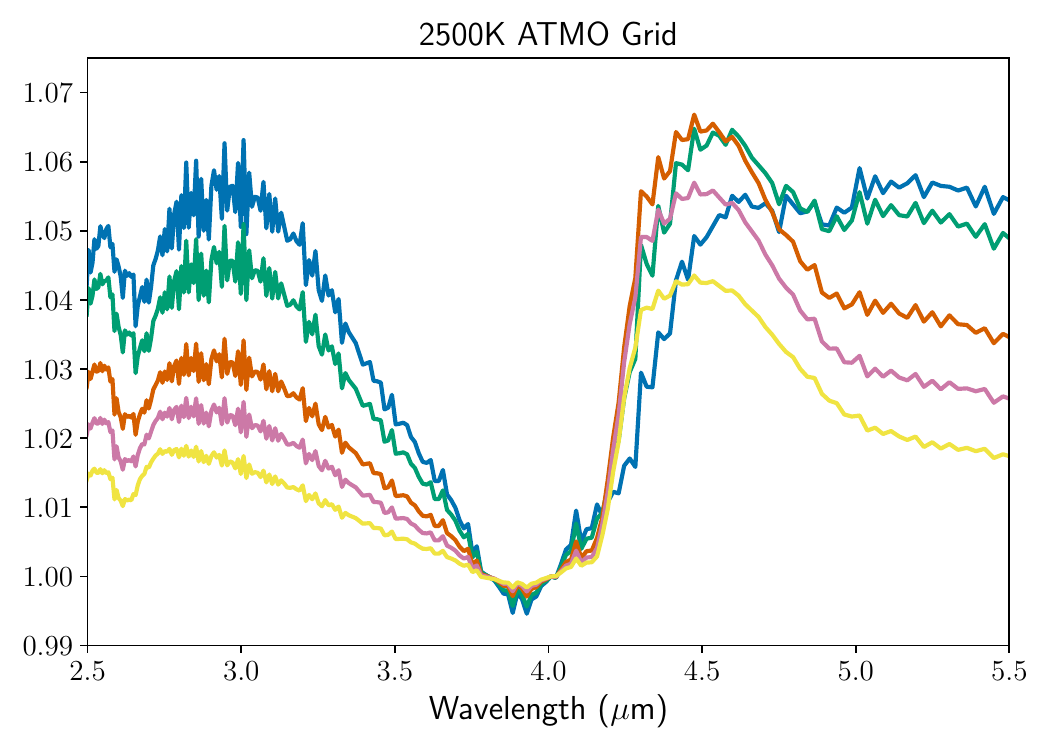}
    \caption{Isothermal transmissions spectra at 1200 (left) and 2500~K (right) at a range of metallicities from the forward models of \cite{goyal:2018}. The spectra are all normalized to 1.0 at 4.0~$\mu$m to compare feature sizes and shapes. While the H$_2$O feature at 1200~K becomes larger with metallicity before decreasing, at 2500~K, the magnitude of H$_2$O monotonically decreases with metallicity.}
    \label{fig:spec_trends}
\end{figure*}

The high metallicity measured in the chemical equilibrium retrievals is surprising. WASP-178b is 1.66$\pm$0.12~$M_{\mathrm{J}}$ \citep[][]{hellier:2019}, so one would not expect the planet to be primordially enriched in heavy elements an order of magnitude more than Jupiter, no matter how the planet formed. The spectrum as shown in Figure~\ref{fig:spec_fits} does not intuitively indicate that there is a lot of H$_2$O in the atmosphere, i.e., the H$_2$O feature from 2.8-3.7~$\mu$m is relatively flat. Considering H$_2$O is a canonical indicator of metallicity \citep{deming:2013}, this behavior is at first puzzling. 

To explain the high-metallicity preference in the retrievals, we compared spectra from the \cite{goyal:2018} grid of \texttt{ATMO} forward models at a variety of $T_\mathrm{iso}$, [Fe/H]\footnote{In reference to the \cite{goyal:2018} models, [Fe/H] refers to the scaling of all heavy elements in the atmosphere, including C and O, not just Fe.}, and C/O. We show these models spread in [Fe/H], normalized at approximately 3.8~$\mu$m, at both an isothermal temperature of 1200 and 2500 K in Figure~\ref{fig:spec_trends}. 

At 1200~K, the spectra behave more intuitively, where increasing the metallicity causes the H$_2$O absorption at 3~$\mu$m to appear larger until about 50$\times$ solar ($\rm [Fe/H] = 1.7$), when the increasing mean molecular weight of the atmosphere starts to decrease the scale height of the atmosphere, causing the spectral features to appear smaller at the highest metallicities. However, at 2500~K, the H$_2$O feature \textit{monotonically} decreases in size with increasing metallicity, seeming to suggest that the more H$_2$O there is in an ultra-hot Jupiter atmosphere, the smaller the spectral feature. One reason for this behavior is the fact that increasing the metallicity raises the photosphere to a level where the H$_2$O abundance is increasingly affected by thermal dissociation, resulting in a correspondingly weaker spectral feature.



The retrieval attempts to match the relatively flat 2.8-3.7~$\mu$m region by \textit{increasing} [O/H]. At high metallicity, however, CO$_2$ starts becoming a significant absorber, even at these high temperatures, though the JWST data show no such prominent CO$_2$ feature. The retrieval can reduce the abundance of CO$_2$ by either increasing the temperature, such that the more stable CO molecule remains the dominant carbon-bearing species, or by increasing the C/O ratio, such that there are fewer O atoms and CO remains the preferred carbon-bearing species. 
The retrievals appear to prefer the former scenario, perhaps to preserve the correct H$_2$O/CO ratio, as \texttt{ATMO} and \pRT{} both find an inverted temperature profile, which will simultaneously dissociate H$_2$O to provide a flat 2.8-3.7~$\mu$m region, while preventing high abundances of CO$_2$. Such an inversion is also predicted by the 1D PHOENIX RCE model. We note that when we mask the 4.1-4.5~$\mu$m region, we retrieve a similarly high [Z/H] of 1.61$\pm$0.28, indicating that the high metallicity constraint is not driven by the presence or absence of CO$_2$ as has been seen for several other JWST datasets to date \citep[e.g.,][]{rustamkulov:2022,alderson:2022}.

As a test, we also ran a chemical equilibrium retrieval with \pRT{} where we fixed the metallicity to solar, a reasonable assumption given the planet's mass. By fixing the metallicity, we can see whether there is an alternative way for the retrieval to fit the data besides moving to higher metallicity. Indeed, a decent goodness-of-fit can be achieved, with the fixed-metallicity retrieval reaching  $\chi^2/N_{\mathrm{data}}$ = 1.07. In order to fit the spectrum, the retrieval finds two modes: a high-temperature ($\approx$2900~K), low-C/O ($<$0.2) mode and a low-temperature ($\approx$1500~K), high C/O ($\approx$0.9) mode, with the best fit being in the latter mode. Both modes can be interpreted as different ways at lowering the magnitude of the H$_2$O bump between 2.8-3.7~$\mu$m, either through thermal dissociation or a lower H$_2$O/CO ratio.

\begin{deluxetable*}{l||c|c|cc||c|c|c}
\tablecaption{Abundance Constraints for WASP-178b from \texttt{ATMO} and \pRT{} Retrievals \label{table:retrieval}}
\tablehead{
\colhead{Retrieval} & 
\colhead{\texttt{ATMO}} & 
\colhead{\texttt{pRT}} & 
\colhead{\texttt{pRT}} & 
\colhead{\texttt{pRT}} & 
\colhead{\texttt{pRT}} & 
\colhead{\texttt{pRT}} & 
\colhead{\texttt{pRT}} \\
\colhead{Chemistry} & 
\colhead{Chem. Eq.} & 
\colhead{Chem. Eq.} & 
\colhead{Chem. Eq.} & 
\colhead{Chem. Eq.} &
\colhead{Free Chem.} & 
\colhead{Free Chem.} & 
\colhead{Free Chem.} \\
\colhead{Dataset} & 
\colhead{JWST-only} & 
\colhead{JWST-only} & 
\colhead{JWST+HST}  &
\colhead{JWST+HST} &
\colhead{JWST-only} & 
\colhead{JWST+HST}  &
\colhead{JWST+HST} \\
\colhead{TP} & 
\colhead{5TP} & 
\colhead{3TP} & 
\colhead{Isothermal}  &
\colhead{3TP}  &
\colhead{Isothermal} & 
\colhead{Isothermal} & 
\colhead{3TP}
}
\startdata
ln\,$Z$& 2026.0 $\pm$ 0.5 &  2559.9 $\pm$0.1 & 3187.6 ± 0.2 & 3193.9 ± 0.2   & 2551.7 ± 0.1  & 3187.8 ± 0.2 & 3190.6 ± 0.2 \\
Weight& - &  - & 0.0018 & 0.9618  & - & 0.0020 & 0.0344 \\
\tableline
H$_2$O$^{\dagger}$ & $-3.30^{+0.80}_{-0.96}$ &  $-3.42^{+0.51}_{-0.65}$ & $-3.30^{+0.47}_{-0.56}$ & $-5.34^{+0.83}_{-0.60}$  & $-5.40^{+0.82}_{-0.82}$ & $-6.26^{+0.14}_{-0.14}$ & $-6.22^{+0.12}_{-0.12}$ \\
CO$^{\dagger}$ & $-2.05^{+0.23}_{-0.26}$ &   $-3.17^{+0.65}_{-0.75}$ & $-5.16^{+0.22}_{-0.34}$ & $-5.61^{+0.20}_{-0.19}$ & $-3.53^{+1.20}_{-1.20}$ & $-5.44^{+0.23}_{-0.24}$  & $-5.38^{+0.20}_{-0.20}$ \\
CO$_2$$^{\dagger}$ & $-5.43^{+0.67}_{-0.83}$ &   $-6.58^{+0.72}_{-0.83}$  & $-7.96^{+0.63}_{-0.70}$ & $-9.95^{+0.55}_{-0.75}$ &  $-7.72^{+0.65}_{-0.65}$ & $-9.58^{+0.99}_{-1.09}$ & $-9.55^{+0.90}_{-1.01}$ \\
SiO$^{\dagger}$ & $-3.55^{+0.65}_{-\infty}$ &   $-3.81^{+0.67}_{-0.68}$ & $-4.62^{+0.15}_{-0.21}$ & $-4.93^{+0.20}_{-0.24}$ & $-7.43^{+2.40}_{-2.50}$ & $-5.08^{+0.18}_{-0.17}$ & $-5.05^{+0.15}_{-0.14}$ \\
Fe$^{\dagger}$ & -- & --& $-4.74^{+0.01}_{-0.01}$ & $-4.69^{+0.11}_{-0.07}$  & -- & $-6.45^{+3.37}_{-3.28}$ & $-6.51^{+3.10}_{-2.95}$ \\
Fe+$^{\dagger}$ & -- & --& $-4.97^{+0.02}_{-0.01}$ & $-5.08^{+0.12}_{-0.45}$ & -- & $-10.06^{+0.93}_{-0.88}$  & $-9.98^{+0.87}_{-0.83}$ \\
FeH$^{\dagger}$ & -- & --& $-9.64^{+0.2}_{-0.03}$ & $-8.77^{+0.37}_{-0.16}$ & -- & $-6.40^{+3.33}_{-3.40}$ & $-6.56^{+3.08}_{-2.92}$ \\
Mg$^{\dagger}$ & -- & --& $-4.45^{+0.01}_{-0.01}$ & $-4.44^{+0.014}_{-0.01}$$^{*}$  & -- & $-6.00^{+3.35}_{-3.35}$ & $-6.04^{+3.00}_{-3.11}$\\
Mg+$^{\dagger}$ & -- & --& --& -- & -- &  $-3.35^{+0.17}_{-0.17}$ &  $-3.45^{+0.21}_{-0.26}$ \\
TiO$^{\dagger}$ & -- & --& $-7.80^{+0.01}_{-0.02}$ & $-8.01^{+0.38}_{-0.55}$& -- & $-10.73^{+0.58}_{-0.48}$ & $-10.72^{+0.52}_{-0.44}$  \\
VO$^{\dagger}$ & -- & --& $-9.35^{+0.01}_{-0.06}$ & $-10.14^{+0.47}_{-0.63}$$^{*}$ & -- &  $-10.29^{+0.83}_{-0.79}$ &  $-10.29^{+0.77}_{-0.73}$\\
\tableline
$[$O/H$]^{\ddagger}$ & $1.64^{+0.16}_{-0.19}$ &$1.56^{+\infty}_{-0.41}$ & $0.4^{+0.45}_{-0.43}$ & $1.49^{+\infty}_{-1.01}$ & $-0.10^{+1.03}_{-1.25}$ &  $-1.60^{+0.16}_{-0.16}$ &  $-1.56^{+0.14}_{-0.14}$ \\
$[$C/H$]^{\ddagger}$ & $1.47^{+0.21}_{-0.25}$  &$0.67^{+\infty}_{-0.74}$ & $-1.61^{+0.25}_{-0.29}$ & $-2.04^{+0.20}_{-0.21}$ &  $0.09^{+1.03}_{-1.25}$ &  $-1.92^{+0.24}_{-0.23}$ &  $-1.86^{+0.20}_{-0.20}$ \\
$[$Si/H$]^{\ddagger}$ & $1.24^{+0.42}_{-0.91}$ & $1.04^{+0.54}_{-0.51}$ & $0.22^{+0.13}_{-0.14}$ & $-0.13^{+0.13}_{-0.12}$ & $-2.81^{+2.19}_{-2.56}$ &  $-0.59^{+0.17}_{-0.17}$  &  $-0.56^{+0.15}_{-0.15}$ \\
$[$Fe/H$]^{\ddagger}$ & -- & --& $0.45^{+0.53}_{-0.65}$ & $-2.51^{+1.20}_{-\infty}$ & -- & $-1.86^{+3.33}_{-\infty}$ & $-2.02^{+3.08}_{-\infty}$ \\
$[$Ti/H$]^{\ddagger}$ & -- & --& $-2.81^{+0.86}_{-\infty}$ & $-2.93^{+0.80}_{-\infty}$ & --& $-3.64^{+0.58}_{-\infty}$ & $-3.63^{+0.52}_{-\infty}$ \\
\tableline
C/O$^{\ddagger}$ & $0.4\pm0.19$ & $0.09^{+\infty}_{-\infty}$ & $0.01^{+0.01}_{-0.00}$ & $0.01^{+0.01}_{-\infty}$ & 0.93$^{+0.65}_{-0.03}$ &  0.30 $^{+0.10}_{-0.10}$  &  0.31$^{+0.09}_{-0.09}$ \\
$[$Si/O$]^{\ddagger}$ & $-0.40^{+0.51}_{-1.00}$  & $-0.44^{+0.74}_{-\infty}$ &  $-0.18^{+0.49}_{-0.49}$ & $-1.64^{+1.06}_{-\infty}$ &  $-2.70^{+2.46}_{-2.56}$ & $1.01^{+0.07}_{-0.07}$ & $1.01^{+0.06}_{-0.06}$ \\
$[$Si/C$]^{\ddagger}$ & $-0.23^{+0.52}_{-1.01}$  & $0.51^{+0.49}_{-0.53}$ &  $1.83^{+0.33}_{-0.30}$ & $ 1.90^{+0.27}_{-0.25}$ &  $-2.90^{+2.44}_{-2.58}$ &  $1.33^{+0.23}_{-0.23}$ &  $1.30^{+0.19}_{-0.19}$ \\
\enddata
\tablenotetext{\dagger}{Quoted abundances from the chemical equilibrium retrievals are the contribution-function-weighted volume mixing ratio for each atom and molecule.}
\tablenotetext{\ddagger}{Quoted abundances from the free retrievals are calculated manually from the posteriors of the atomic and molecular abundances.}
\tablenotetext{*}{Abundance not varied by a free parameter, but representative of solar abundance at the retrieved temperature and pressures.}
\end{deluxetable*}

\subsubsection{JWST-Only Free Chemistry Retrievals}\label{sec:results:jwstonly:free}

In addition to the chemical equilibrium retrievals described above, we also ran retrievals with \pRT{} where the abundance (via the mass fraction) of individual chemical species were treated as free parameters and the abundances are considered uniform with pressure. Figure~\ref{fig:spec_fits} shows the best-fit spectrum from the JWST-only free retrieval, compared with those from the chemical equilibrium retrievals described above. The free retrieval reaches $\chi^2/N_{\mathrm{data}}$ = 1.03. This is actually somewhat worse than the chemical equilibrium retrievals. This could be a sign that thermal dissociation is changing the molecular abundances across the transit photosphere, which is not accounted for when we keep abundances uniform with pressure. Even if the free retrieval can adjust to the abundance at the photospheric pressure, if the abundance actually varies appreciably across the pressures probed by our observations, this subtle effect would not be accounted for in the free retrieval. 

 As above, the retrieved VMRs are shown in Figure~\ref{fig:vmrs}. Note that because \pRT{} retrieves abundances in mass fraction, rather than volume mixing ratio, a constant mass fraction with pressure will lead to a bend in the volume mixing ratio (VMR) where the thermal dissociation of H$_2$ into H begins, since this will nearly double the number of particles in the atmosphere. 
The measured abundances listed in Table~\ref{table:retrieval} are then the VMR at a mean molecular weight of 2.3. Since WASP-178b's atmosphere is hot enough that the thermal dissociation of molecules is relevant, one should consider the freely retrieved abundance to be representative of the transmission photosphere (i.e., the pressure level probed by the observations), and not necessarily the atmosphere or planet as a whole.

Interestingly, the abundances measured in the free retrieval tell a different story compared to the chemical equilibrium retrievals. As shown in Table~\ref{table:retrieval} and Figure~\ref{fig:abund_graphic}, relatively low abundances are measured in the free retrievals. This can at least in part be explained by thermal dissociation: the free retrieval will be measuring abundances at the photosphere, which are likely undergoing at least some amount of thermal dissociation. Figure~\ref{fig:vmrs} shows that in the chemical equilibrium case, the abundance of H$_2$O starts dissociating around 1 mbar, reaching the abundances measured by the free retrieval at about 0.01-0.1 mbar. However, CO and SiO also undergo dissociation, but at an altitude above the transit photosphere. 

While H$_2$O, CO, and CO$_2$ all exhibit bounded constraints, SiO is not well measured, implying that IR absorption of SiO is not detected, despite the large NUV absorption attributed to SiO. We explore this further in the combined JWST+HST retrievals in Section~\ref{sec:results:full_rets} and compare to WASP-121b, where SiO has been detected in the IR, in Section~\ref{sec:dis:sioIR}.

\subsection{Full UV-to-IR Retrievals}\label{sec:results:full_rets}

It is clear that despite the definite molecular detections in the JWST/NIRSpec/G395H transmission spectrum of WASP-178b, the IR data alone is insufficient to place robust constraints on the planet's composition. To that end, we combine the JWST transmission spectrum with observations from all of the spectroscopic grisms of HST/WFC3. The combination of G280, G102, and G141 provides nearly continuous spectral converage from 0.2-1.7~$\mu$m, probing the NUV to the NIR. For the full spectral range, we use \pRT{}, which can model both the refractory species present at short wavelengths and the volatile species at longer wavelengths. 

Similar to the JWST-only data in Section~\ref{sec:results:jwstonly}, we perform both chemical equilibrium and free retrievals. In chemical equilibrium, we retrieval additional elemental abundances, adding [Fe/H] and [Ti/H] to account for species like Fe, Fe+, FeH, and TiO that may absorb at shorter wavelengths. Similarly, we added Fe, Fe+, FeH, Mg, Mg+, TiO, and VO to the free retrieval. We also explore how the retrieval results depend on the temperature structure parametrization, since a temperature inversion has been hypothesized to help explain the large absorption in the NUV \citepalias{lothringer:2022}.

We note here an interesting convergence in the retrieved TP profiles, seen in the bottom panel of Figure~\ref{fig:TP}. While the JWST-only retrievals retrieved profiles that varied by $>1000$~K, the full-spectrum retrievals all converge towards a temperature of about 3200~K at 1 mbar, about where we would expect the transit photosphere to be. This also agrees well with the 1D self-consistent model, which assumes full-heat redistribution. It appears that by adding the short-wavelength data, likely the large-amplitude features from the NUV observations especially, the retrievals all prefer a hotter atmosphere scenario, regardless of the other retrieval assumptions. 

The retrieved temperature between the retrievals of about 3200~K at 1 mbar is significantly warmer than the planet's equilibrium temperature of $\sim2450$~K. it could be the case that our observations are probing a strong temperature inversion, as expected from the self-consistent model. The temperature may be further biased towards hot temperatures because of the puffy, strongly-irradiated dayside \citep{pluriel:2020a}. It is unclear why the chemical equilibrium retrieval prefers a non-inverted atmosphere, while the free retrieval finds an inverted atmosphere. We hypothesize that since the UV refractory species probe higher in the atmospheres compared to the molecular volatile species, the chemical equilibrium retrieval may be using a non-inverted profile to refine the scale height or absorption between these two regimes.



\subsubsection{HST+JWST Chemical Equilibrium Retrievals}\label{sec:results:full_rets:chemeq}

The best-fit full-spectrum chemical equilibrium retrieved model results in a $\chi^2/N$ of 1.169 and 1.177 for the isothermal and 3-parameter TP (3TP) retrievals, respectively. While somewhat worse than the JWST-only fits, the best fit shown in Figure~\ref{fig:spec_fits} is able to match the very large absorption in the NUV, the flat optical, and the muted H$_2$O and CO in the infrared.

\begin{figure*}
    \centering
    \includegraphics[width=1.0\linewidth]{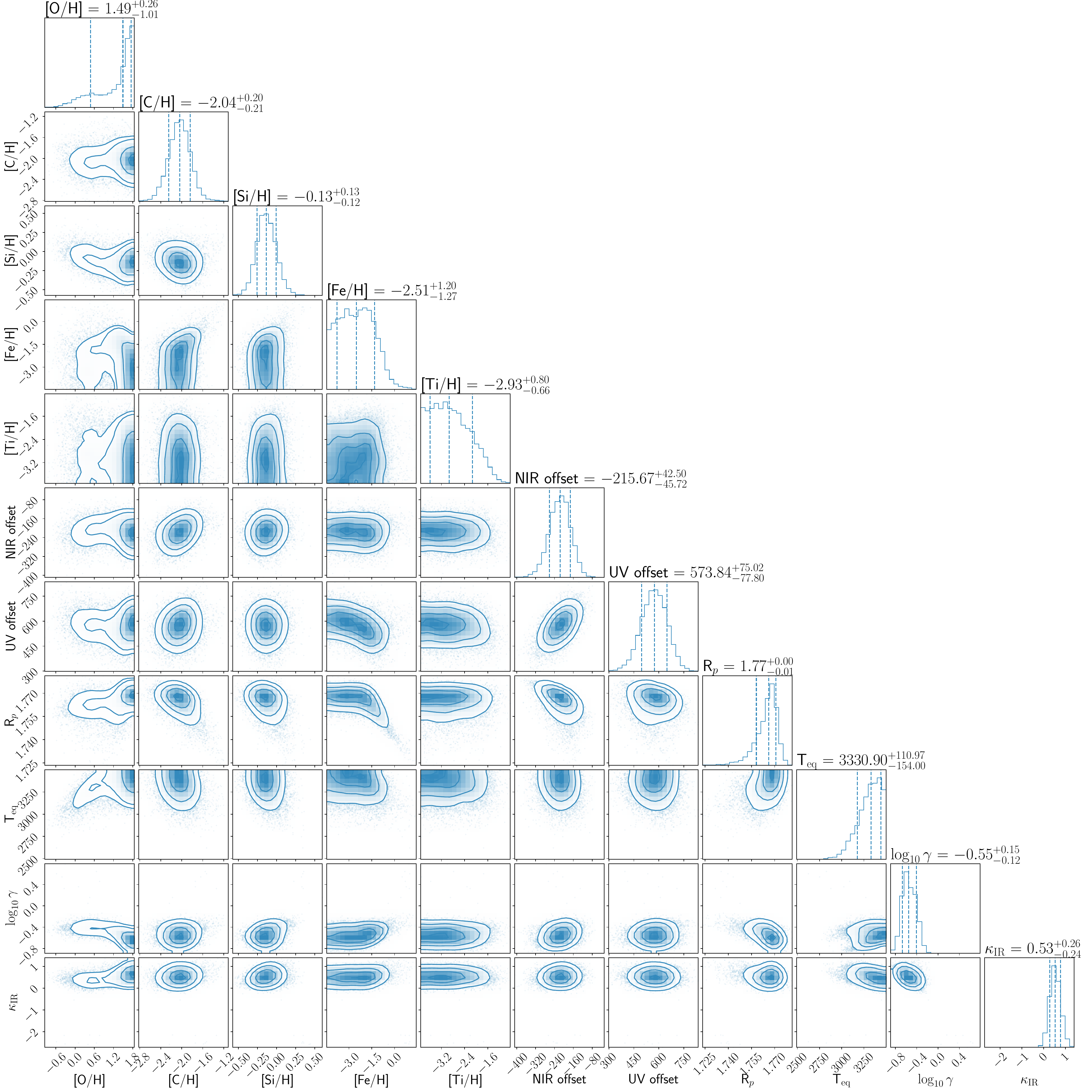}
    \caption{Retrieved posteriors for the full-spectrum HST+JWST retrieval under chemical equilibrium and with the 3-parameter temperature structure from \cite{guillot:2010}.\added{ Elemental ratios are represented relative to solar, data offsets are in ppm, the planet radius $R_\mathrm{p}$ is in units of Jupiter radii, the temperature is in Kelvin, $\gamma$ is unitless, and $\kappa_\mathrm{IR}$ is in cm$^2$/g.}}
    \label{fig:full_chemeq_corner_pg}
\end{figure*}

\begin{figure*}
    \centering
    \includegraphics[width=1.0\linewidth]{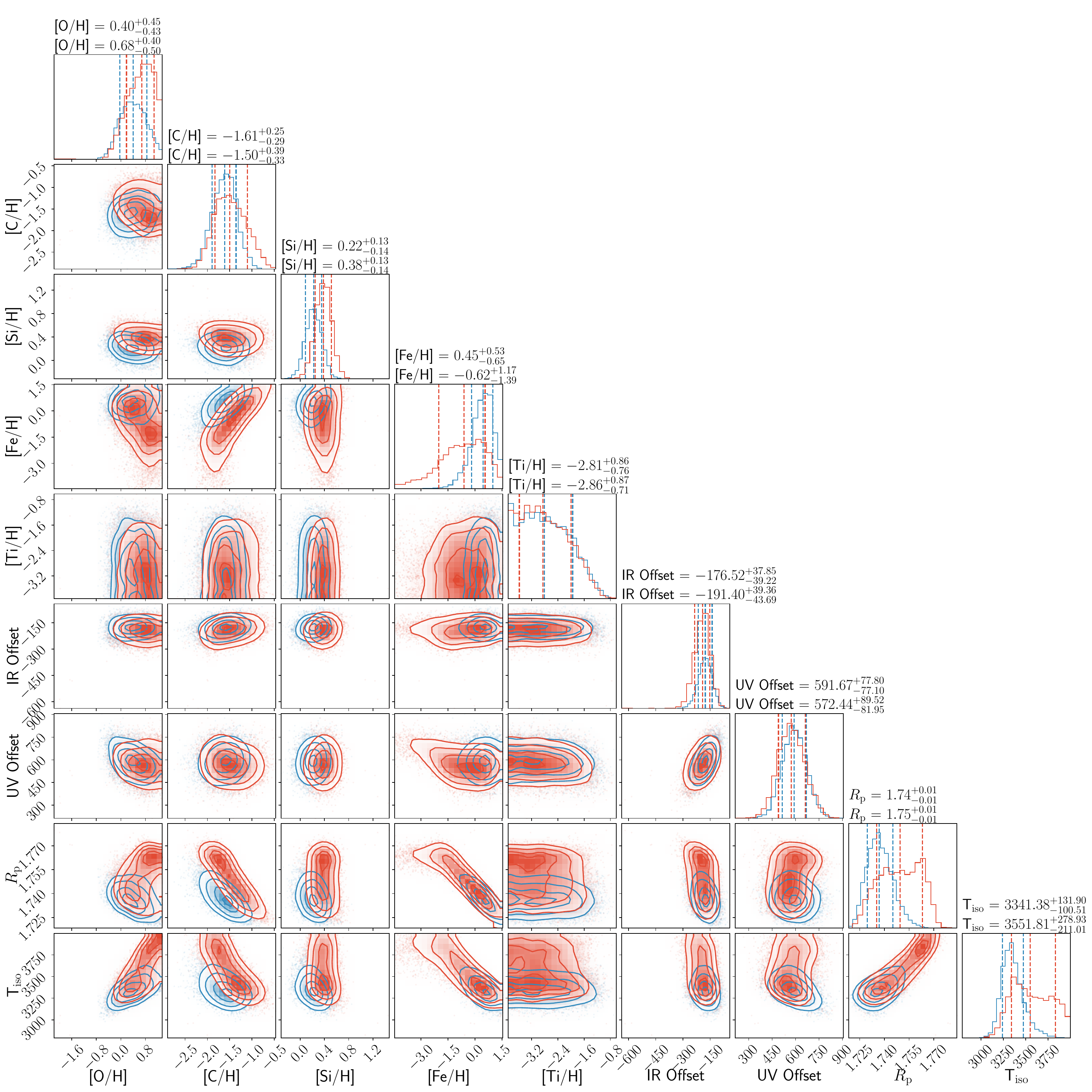}
    \caption{Retrieved posteriors for the full-spectrum HST+JWST isothermal chemical equilibrium retrieval using both the \texttt{FIREFLy} (blue, top titles) and \texttt{transitspectroscopy} (red, bottom titles) reductions.}
    \label{fig:full_chemeq_corner}
\end{figure*}

As in Section~\ref{sec:results:jwstonly}, the molecular abundances retrieved are shown in Table~\ref{table:retrieval} and Figure~\ref{fig:abund_graphic} with vertical profiles of H$_2$O, CO, CO$_2$, and SiO shown in Figure~\ref{fig:vmrs}. We also show the posterior cross-sections for the 3TP and isothermal retrievals shown in Figures~\ref{fig:full_chemeq_corner_pg} and \ref{fig:full_chemeq_corner}, respectively. While the quality of fits for the isothermal and 3TP retrievals are similar, the agreement between elemental abundances varies. Constraints on [C/H], [Si/H], and [Ti/H] are similar between the two retrievals, while [O/H] and [Fe/H] are significantly different. We discuss each measurement in turn. 

\paragraph{[O/H]}
[O/H] is found to be consistent with solar within 1$\sigma$ for the isothermal retrieval, while the 3TP retrieval prefers a super-solar [O/H] reaching the upper prior, similar to what was found in the JWST-only retrievals from Section~\ref{sec:results:jwstonly}. The 3TP retrieval does, however, allow for a long tail towards lower [O/H], overlapping with the isothermal [O/H] measurement. The 3TP [O/H] measurement appears to be negatively correlated with $\gamma$ (see Figure~\ref{fig:full_chemeq_corner_pg}), the ratio between the optical and infrared mean opacity, which determines the magnitude of any non-isothermal behavior (i.e., the inversion). This correlation indicates that the retrieval is also finding a more isothermal mode approaching solar [O/H] matching the isothermal retrieval's behavior.

This is reflective of the general degeneracy in ultra-hot Jupiter retrievals, whereby the [O/H] and temperature can be effectively traded-off due to the effects of thermal dissociation. For example, a measurement of approximately solar H$_2$O at the transit photosphere could indicate either a relatively cool, solar [O/H] atmosphere or a hotter, super-solar [O/H] atmosphere undergoing thermal dissociation of molecules. Measuring CO in addition to H$_2$O does not fix this problem, as the [C/H] or C/O ratio can also be adjusted.

\paragraph{[C/H] \& C/O}
Both retrievals agree on significantly sub-solar [C/H] abundances depleted by a factor of between 10-100, with C/O ratios of $<0.02$ at 3$\sigma$. It appears that the large-scale height implied by the NUV absorption requires a low [C/H] to explain the relatively muted CO feature in the IR. A very low [C/H] would have important implications for planet formation \citep{oberg:2011,mordasini:2016}. We discuss these implications in Section~\ref{sec:formation}. 

\paragraph{[Si/H]}
Both retrievals also agree on a [Si/H] abundance close to solar. This is interesting because SiO was not clearly detected in the IR, yet the retrieval uses SiO to fit the NUV absorption feature. This suggests that the non-detection of SiO in the IR does not preclude significant absorption in the NUV, where the SiO opacity is stronger by about four orders of magnitude \citep{yurchenko:2018}. [Si/H] is also the most robustly constrained refractory abundance for this dataset, so we interpret its significance relative to planet formation in Section~\ref{sec:formation}.

We do note a caveat in our measure of [Si/H]. Because much lower pressures are probed by SiO in the UV ($10^{-3} - 10^{-6}$ bar, see Figure~\ref{fig:vmrs}), we may be reaching part of the atmosphere that is not strictly in hydrostatic equilibrium. In this case, our abundance measurements would be an overestimate, as the absorption is spread along a longer atmospheric column than our hydrostatic models assume.

\paragraph{[Ti/H]}
Again, both retrievals agree on a significantly sub-solar [Ti/H] abundance due to the apparent absence of TiO absorption at optical wavelengths. The depletion of Ti relative to other refractories has been seen in high-resolution studies of ultra-hot Jupiters \citep{pelletier:2023,gandhi:2023} and points to cold-trapping of Ti on the nightside of these planets \citep{parmentier:2013,beatty:2017a,hoeijmakers:2024}.

\paragraph{[Fe/H]}
Lastly, the retrievals differ in the measurement of [Fe/H]. The 3TP retrieval places very wide constraints centered at very low abundance, while the isothermal retrieval prefers near-solar [Fe/H] within 1$\sigma$. The difference in [Fe/H] is responsible for the visible difference between the chemical equilibrium best-fits in Figure~\ref{fig:spec_fits}, where the isothermal fit exhibits neutral Fe absorption between 0.3-0.45~$\mu$m that the 3TP retrieval discards by depleting [Fe/H]. It is not clear why the isothermal retrieval keeps a higher [Fe/H], but we hypothesize that it is in order to help fit the NUV spectrum with increased metal absorption due to the isothermal retrievals inability to increase the scale height through a temperature inversion.

\paragraph{[Si/O] \& [Si/C]}
We can use pairs of elemental abundances relative to H to get relative abundance ratios that can sometimes be better constrained than any absolute ratio \citep{gibson:2022}. For example, [Si/O] is calculated to be consistent with solar in both the isothermal and 3TP full-spectrum chemical equilibrium retrievals, though with uncertainty up to a dex. \added{If [O/H] is even greater than our prior, the corresponding [Si/O] would be lower.} [Si/C] is found to be consistently and significantly super-solar. Each of these ratios can be used as independently constraining refractory-to-volatile ratios \citep[][]{lothringer:2021}.

\medskip
We also use the full HST+JWST spectrum to test the retrieval's sensitivity to the JWST data reduction. Running the same isothermal chemical equilibrium retrieval with the JWST/NIRSpec data reduction from \texttt{transitspectroscopy} results in consistent atmospheric constraints, as shown by the posteriors in Figure~\ref{fig:full_chemeq_corner}.

\subsubsection{HST+JWST Free Chemistry Retrievals}\label{sec:results:full_rets:free}

Finally, we computed retrievals assuming free chemistry as we did in Section~\ref{sec:results:jwstonly}. As shown in Figure~\ref{fig:spec_fits}, the best-fit free chemistry retrieval to the full HST+JWST dataset has $\chi^2/N$ of 1.170, only marginally better than the equilibrium chemistry fit at $\chi^2/N=1.177$. Full posterior distribution corner plots are shown in Appendix~\ref{appendix:corners} Figures~\ref{fig:full_free_corner} and \ref{fig:full_free_corner_PG}.

We find good agreement between the isothermal and 3TP retrievals when freely retrieving abundances. This suggests that the differences between the temperature structure retrievals in Section~\ref{sec:results:full_rets:chemeq} were not primarily due to fitting the spectral features differently (besides in the case of the Fe absorption discussed above), but rather through the extrapolation of the measured photospheric abundances to bulk abundances through chemical equilibrium assumptions. This is further reflected in the fact that the molecular and atomic abundances (weighted by the contribution function) in the chemical equilibrium retrievals listed in Table~\ref{table:retrieval} and Figure~\ref{fig:abund_graphic} show consistent abundances between the isothermal and 3TP retrievals, yet the extrapolated elemental abundances vary. The one exception is H$_2$O, where the 3TP chemical equilibrium retrieval found a lower H$_2$O abundance than the isothermal chemical equilibrium retrieval in the photospheric region, yet a higher [O/H] elemental abundance, highlighting the important role of H$_2$O dissociation in measuring ultra-hot Jupiter compositions. 

Additionally, the atomic and molecular abundances measured in the HST+JWST free retrievals agrees broadly with those found in chemical equilibrium. The difference, however, is in the extrapolated elemental abundances. In all cases, the free-retrieval elemental abundances are found to be significantly sub-solar. This is expected and simply due to the fact that thermal dissociation is not accounted for in the free retrieval when extrapolating the measured atomic and molecular abundances to their elemental abundances. Hence, the free chemistry retrievals are likely to underestimate the true elemental abundances.

\section{Discussion}\label{sec:discussion}

\subsection{Bayesian Model Comparison}

In this study, we have tested a number of retrieval models in an attempt to measure the composition of WASP-178b. To determine which model assumptions or parametrization are justifiable, here we compare the Bayesian Evidence as calculated in the Nested Sampling retrievals. The ratio between the evidences (or equivalently, the difference in their logarithms) is the Bayes factor and can be used to compare the statistical support for one model over another \citep[e.g.,][]{trotta:2008}.

Table~\ref{table:retrieval} lists the evidence for the various retrieval scenarios, including free versus equilibrium chemistry and the isothermal versus 3-parameter temperature structure (3TP) parametrization. We note that the priors used in the retrievals are listed in Table~\ref{tab:priors} and that the computed evidence can depended arbitrarily on the chosen priors.

In general, for both the JWST-only and JWST+HST retrievals, we find that the 3TP parametrization is justified at the $3\sigma$ level over an isothermal approximation, with Bayes factors of $\gtrapprox 3$. Similarly, chemical equilibrium is preferred over free chemistry, likely due to its ability to fit the atmosphere adequately with fewer free parameters. Despite these results, we still recommend fitting the data with a variety of assumptions to evaluate the robustness of any single retrieval result. We note that some free parameters, like the FeH or VO abundance, could likely be removed from the free retrieval with little consequence to the fit, thus improving the evidence.

With the evidence for each retrieval, we can calculate a weight for each retrieval to combine the posteriors into a final inferred constraint on the elemental abundances. We first define a normalized evidence,

\begin{equation}
    \Delta \ln{Z_i} = \ln{Z_i} - \max{\left(\ln{Z}\right)},
\end{equation}

\noindent before calculating the weight as

\begin{equation}
    w_i = \frac{e^{\Delta \ln{Z_i}}}{\sum_j e^{\Delta \ln{Z_j}}}
\end{equation}

\noindent where $Z$ is the Bayesian Evidence from the retrievals and $i$ and $j$ refer to the different retrieval models \citep{hoeting:1999}. The resulting weights are listed in Table~\ref{table:retrieval}, with the 3TP chemical equilibrium retrieval clearly dominating the weight. We then combine the posterior samples from all \pRT{} JWST+HST retrievals by taking a proportional number of samples from each retrieval's posterior to create a 7000 sample weighted-average posterior. The results are listed in Table~\ref{tab:weighted_avg} and shown in Figure \ref{fig:abund_graphic}. The evidence-weighted abundances are dominated by those measured in the 3TP chemical equilibrium retrieval, which had the highest evidence and therefore highest weight.

\begin{table}
\centering
\caption{Final Evidence-Weighted Average Abundance Measurements for WASP-178b from HST+JWST Retrievals\label{tab:weighted_avg}}
\begin{tabular}{lr}
\hline
Parameter & Value \\
\hline
$[$O/H$]$ & $1.47^{+0.28}_{-1.10}$ \\
$[$C/H$]$ & $-2.03^{+0.20}_{-0.21}$ \\
$[$Si/H$]$ & $-0.14^{+0.13}_{-0.14}$ \\
$[$Fe/H$]$ & \replaced{$-2.45^{+1.23}_{-1.31}$}{$-2.45^{+1.23}_{-\infty}$} \\
$[$Ti/H$]$ & \replaced{$-2.95^{+0.81}_{-0.81}$}{$-2.95^{+0.81}_{-\infty}$} \\
$[$Si/O$]$ & $-1.56^{+0.97}_{-0.30}$ \\
$[$Si/C$]$ & $1.90^{+0.25}_{-0.25}$ \\
C/O & $0.00^{+0.01}_{-0.00}$ \\
\hline
\end{tabular}
\end{table}

\subsection{Transmission Spectra of Ultra-hot Jupiters}

The complication in measuring a consistent and robust atmospheric composition described in the retrieval results above highlight the complexity of interpreting observations of ultra-hot Jupiter atmospheres. On the face of it, the NIRSpec/G395H transmission spectrum of WASP-178b appears to be straightforward: somewhat weak but clearly apparent H$_2$O and CO are detected at relatively good S/N, with no identifiable unexpected features. Assuredly, some of the complications come from thermal dissociation of H$_2$O (and to some degree, H$_2$). The chemical equilibrium retrievals of WASP-178b are generally forced to higher [O/H] to account for this dissociation. There can thus be a degeneracy between the temperature structure and the molecular abundances. 

Ultra-hot Jupiters are subject to a number of potentially relevant 2- and 3D effects. The strong day-night temperature gradient means that there are contributions from both the day- and night-side hemispheres of the planet at the terminator \citep[][]{pluriel:2020a}. The puffy, hot, and dry dayside results in a stronger CO spectral feature, while the H$_2$O remains weak because its only contribution is coming from the colder, more compact nightside hemisphere. Vertical and east-west effects can also be relevant, as pointed out by \citep{pluriel:2022}.


Unfortunately for the sake of disentangling all of these effects, the 1D retrievals presented here fit the data quite well. We obtain $\chi^2$ per data point $<1.2$ in our relatively simple retrievals across the entire 0.2-5.1$\mu$m range, despite probing around 3 dex of pressure, with data from two different telescopes, with four different observing modes, on four different detectors, at four different epochs, measuring atoms, ions, and molecules. This means that we may not be able to use Bayesian model comparison to justify a more complex model. However, since we know these atmosphere to not be 1D or isothermal, physically-motivated choices may need to be made for datasets like this, even if they are not statistically justified. We will explore these more complex models in future work.

Another avenue towards obtaining more robust elemental abundances for atmospheres such as WASP-178b's is to search for other carbon and oxygen-bearing species that can help trace dissociation. This would include the products of H$_2$O dissociation, OH and atomic O. To that end, we tested exploratory free retrievals with OH included, and did not find significant constraints from our dataset. OH and O may be probed best at high-resolution, which when combined with low-resolution observations, may provide the most complete picture of atmospheric composition.

\subsection{Comparison to WASP-121b}

WASP-121b is another ultra-hot Jupiter that has also been observed in transit with JWST/NIRSpec/G395H \citep{gapp:2025}as part of the phase curve observation of Program 1729 (PIs: Evans-Soma, Kataria, \citealt{mikal-evans:2023}). WASP-121b is about 100~K cooler than WASP-178b and orbits a much later-type star (F6 versus A0).

WASP-178b's infrared spectrum shows some distinct differences with WASP-121b. As shown in Figure~\ref{fig:spec_compare}, WASP-121b's spectrum shows stronger H$_2$O features, discernible SiO absorption at 4.1~$\mu$m, and a divergence from WASP-178b at 4.9~$\mu$m. As mentioned in Section~\ref{sec:results}, WASP-121b shows clear H$_2$O absorption at 1.2 and 1.4~$\mu$m, while WASP-178b remains much more featureless across the optical and infrared. At shorter wavelengths, the two planets agree much more closely, with each showing similarly large NUV absorption as discussed in \cite{evans:2018,sing:2019} and \citepalias{lothringer:2022}. 

\subsubsection{[O/H] and Thermal Dissociation}

As in WASP-178b, H$_2$O dissociation was found to be important for retrievals of WASP-121b. A comparison between the inferred degree of dissociation for WASP-178b and WASP-121b would inform our understanding of the atmospheric physics in this regime since even more H$_2$O could be dissociated in WASP-178b compared to WASP-121b due to its $\sim100$~K warmer equilibrium temperature and the higher UV flux from the earlier-type host star, which can lead to a stronger temperature inversion \citep{lothringer:2019}.

Similar to our results here, chemical equilibrium retrievals of WASP-121b's JWST/NIRSpec spectrum preferred high-metallicity, while free retrievals measured a depleted abundance of H$_2$O. This was interpreted to be a result of thermal dissociation, where the photospheric abundance is depleted, but chemical equilibrium retrievals extrapolate a high-metallicity bulk elemental abundance. As we discussed here, the high metallicity in the chemical equilibrium retrievals can also raise the photosphere to lower pressures, where thermal dissociation is more effecting, thereby muting the molecular features. Two-dimensional day/night retrievals performed for WASP-121b with \texttt{NEMESIS} favored a higher H$_2$O abundance on the nightside than the dayside, consistent with thermal dissociation of H$_2$O and complicating the inference of bulk [O/H] abundances \citep{gapp:2025}.

\subsubsection{C/O Ratio}

A stark difference in WASP-121b and WASP-178b's transmission are the inferred C/O ratio. \cite{gapp:2025} found a significantly super-solar C/O ratio, especially when accounting for the phase curve in the out-of-transit baseline. The high C/O of WASP-121b is backed up with two independent ground-based high-resolution studies finding similarly elevated values \citep{smith:2024,pelletier:2025}. 

For WASP-178b however, we find a much lower C/O ratio, with full-spectrum chemical equilibrium retrievals measuring [C/H] to be 10--100 $\times$ (at 1$\sigma$) depleted relative to solar metallicity. While our shorter transit baseline could affect our transmission spectrum and explain some of the difference between the two planets, this could also signal an important compositional difference between WASP-121b and WASP-178b, which would have implications for the planet's formation (see Section~\ref{sec:formation}).

\subsubsection{Nightside Contamination}

As mentioned above, \cite{gapp:2025} fit WASP-121b's time series light curve as part of a phase curve observation. When the spectroscopic transit depths are calculated, the out-of-transit baseline was treated as 1) a portion of the sinusoidal phase curve and 2) flat, with each method giving slightly different results, shown in Figure~\ref{fig:spec_compare}, attributed to flux contamination from the dayside hemisphere rotating out of view and then into view.

This phenomenon will impact all planetary transits. Unfortunately, our WASP-178b observations do not include sufficient baseline to detect the phase curve behavior, so we must treat the out-of-transit baseline as flat. However, this effect is likely less important for WASP-178b since its period is over twice as long as WASP-121b's, 3.34 days \citep{hellier:2019} versus 1.27 days \citep{delrez:2016}. Thus, the out-of-transit baseline will be flatter during our observations than it would be for WASP-121b.

The nightside of WASP-178b's atmosphere may also contaminate the transit depths in a similar fashion. Theoretically, this could serve to mute the CO absorption feature since the planet-to-star flux ratio generally increases with wavelength. Assuming a 2000~K nightside temperature and stellar parameters as listed in Table~\ref{tab:orb_params}, we can expect nightside contamination of the transmission spectrum  between optical and infrared wavelengths on the order of only 15 ppm \citep{kipping:2010b}, which is much smaller than our measurement uncertainties.

\subsubsection{SiO and Refractory Abundances}\label{sec:dis:sioIR}

\cite{gapp:2025} found evidence of SiO absorption around 4.2~$\mu$m, providing an exciting avenue to measure refractory and volatile abundances from the same transmission spectrum. Free retrievals subsequently measured SiO abundances of about 300 ppm. Absorption by SiO was already suggested from the NUV spectrum of WASP-121b \citepalias{lothringer:2022}. While this same NUV absorption was seen in WASP-178b, the NIRSpec/G395H spectrum presented here shows no obvious absorption by SiO in the infrared. Our JWST-only free retrievals find a $1\sigma$ upper limit of about 20 ppm with full-spectrum free retrievals measuring SiO to be between 6.5 and 12.6 ppm at $1\sigma$ confidence. As with [C/H], this might signal another composition difference between WASP-121b and WASP-178b, with the former being more rich in refractory elements. However, the same caveats relating to [O/H], namely thermal dissociation, and the higher temperature of WASP-178b apply here to the [Si/H] measurement.

While the high-resolution analyses of \cite{smith:2024} and \cite{pelletier:2025} agree on the C/O ratio of WASP-121b, they each come to somewhat different conclusions regarding the refractory-to-volatile ratio, with the former finding a slightly super-stellar refractory-to-volatile ratio and the latter finding evidence for the opposite. The abundances measured with JWST from \cite{gapp:2025} do not clear up the picture either, as these depend on the retrieval and baseline assumptions. Retrieving the complete low-resolution UVOIR spectrum of WASP-121b could clarify the interpretation of the planet's refractory-to-volatile ratio, particularly when combined with recent NIRISS/SOSS observations from Program JWST-GTO-1201 (PI: Lafreni\`ere) to produce a continuous spectrum.

\subsection{Constraints on Formation}\label{sec:formation}

\subsubsection{Clues from System Properties}

The goal of the present study is to measure the refractory-to-volatile ratio of WASP-178b in order to constrain the planet's formation location and mechanism. From our stellar characterization in Section~\ref{sec:stellar}, we now know WASP-178b formed within the past $\sim$100 Myr. This age is well within the timescale for the Kozai-Lidov mechanism to decrease the planet's semi-major axis and leave the planet mis-aligned \citep{fabrycky:2007}. However, this would be on the shorter end of possible timescales for secular interactions with other planets in the system to have enough time to excite the orbit \citep{wu:2011,chontos:2022}. As the star is above the Kraft break of $\gtrapprox$6200 K, the spin of the star and planet's orbit will not re-align \citep{kraft:1967}.

\cite{hellier:2019} noted excess astrometric noise from WASP-178's 5-parameter astrometric solution in Gaia DR2 \citep{gaia:2018}, which persists in DR3 \citep{Gaia:2023} at 0.140 mas. This excess noise is considered significant if \texttt{astrometric\_excess\_noise\_sig} $>$ 2 and this value is 28.9 for WASP-178. This excess noise could be from an unseen companion, perhaps the same that interacted with WASP-178b to send it into its present-day short-period orbit. At 10 AU (0.025$\farcs$ angular separation), this excess noise as astrometric wobble would correspond to a minimum mass of about 12~$M_\mathrm{J}$, while at 100 AU  (0.25$\farcs$), this would correspond to a minimum mass of about 1.2~$M_\mathrm{J}$. Given the age of the system, a planet between 1.2 and 12~$M_\mathrm{J}$ planet would be between about 400 and 1000~K \citep{marley:2021}. Despite the distance to the WASP-178 system, such companions may be detectable with next-generation direct imaging facilities.

We note that with high spatial resolution speckle imaging of WASP-178 with Gemini-South/Zorro, \cite{rodriguezmartinez:2020} ruled out bright companions from 17 mas to 1.7$\farcs$ at a contrast of 4.2 mag at 562 nm and from 28 mas out to 1.7$\farcs$ at a contrast of 5-7 $\Delta$-mag at 832 nm. With the Southern Astrophysical Research Telescope SOAR, only a 7.1 mag source was seen at 2.4$\farcs$ separation in $I_c$-band observations (738--909 nm). However, no similar source is seen in Gaia and at such a large angular separation, this source is unlikely to be gravitationally bound if it is real.

As mentioned in Section~\ref{sec:stellar}, WASP-178 is a likely Am star, with peculiar metal abundances (weak Ca and Sc lines with higher abundances of other heavy metals) and an unusually slow rotation \citep[see Table~\ref{tab:orb_params}][]{hellier:2019,rodriguezmartinez:2020}. In our comparison to open clusters of similar age and distance, we inferred a primordial abundances for the WASP-178 system as listed in Table~\ref{tab:orb_params}, broadly consistent with solar metallicity to within 0.15 dex.


\subsubsection{Clues from WASP-178b's Atmosphere}

Although variations in retrieval constraints complicate unambiguous conclusions about formation, interpreting the similarities across different \added{retrieval} scenarios can still yield robust insights into WASP-178b's formation mechanism. Any [Si/H] measurement can be interpreted as a strict lower-limit on the bulk abundance \added{of Si }since there may be additional Si cold-trapped into nightside clouds, dissociated into atomic Si, or trapped in the planet's core. The lowest [Si/H] value we measured from the full 0.2-5.1~$\mu$m spectrum came from the isothermal free-chemistry retrieval with a $1\sigma$ range between 0.17 and 0.38$\times$ solar. Therefore, the refractory material as traced by Si, can be depleted by no more than a factor of about 10 relative to solar. The largest [Si/H] measurements come from the chemical equilibrium retrievals, which find [Si/H] much closer to solar. This range is consistent with the [Fe/H] measurements from high-resolution ground-based surveys of similar planets in \cite{gandhi:2023}.

Qualitatively, an [Si/H] abundance within an order of magnitude of solar metallicity suggests a formation that was not solely driven by gas-accretion, though this is clear from other ultra-hot Jupiter measurements. On the other hand, a significantly sub-solar C/O ratio would seemingly rule out a distant gas-dominated accretion scenario, suggesting accretion by oxygen-rich gas or planetesimals. Combining both the sub-solar C/O and [Si/H] that is broadly consistent with solar would point towards an oxygen-rich planetesimal/pebble formation scenario. If [O/H] is indeed super-solar in WASP-178b, this would suggest a sub-solar refractory-to-volatile ratio and would further indicate that the solids accreted by the planet were ice-rich in addition to being oxygen-rich, pinpointing atmospheric enrichment occurring between the H$_2$O and CO$_2$ ice line.

Another way to view these constraints is through the [Si/O] and [Si/C] ratios. While the free retrievals find a super-solar [Si/O], the more reliable constraint comes from the full-spectrum chemical equilibrium retrievals, which find a sub-solar to solar [Si/O]. This contrasts strongly with [Si/C], which is universally found to be super-solar in the full-spectrum retrievals, regardless of temperature structure parametrization or chemistry assumptions. These constraints could imply, as above, that WASP-178b was enriched with oxygen-rich, carbon-depleted, refractory-containing material, perhaps some rock-H$_2$O ice mixture. 

This analysis also demonstrates that there is not one single refractory-to-volatile ratio to quote for a given planet, but ratios of multiple species can more powerfully discern the detailed planetary composition. The combination of [Si/O] and [Si/C] measurements helps elucidate the potentially complex behavior of rocky and icy material participating in material enrichment. Even in the presence of significant molecular dissociation, consistent and informative abundance constraints can still be gained.

\section{Conclusion}\label{sec:conclusion}

In this work, we presented the 0.2--5.1\,$\mu$m transmission spectrum of WASP-178b, combining all of HST/WFC3's spectroscopic grisms (0.2--1.7\,$\mu$m) with JWST/NIRSpec/G395H (2.8--5.5\,$\mu$m). The entire UV-to-IR spectrum shows strong refractory absorption in the NUV but muted H$_2$O and CO absorption at longer wavelengths. The spectrum is qualitatively similar to WASP-121b in that there is strong NUV absorption and no CO$_2$, but the relative magnitude of the spectral features is quite different between the two planets.

We fit these data with a suite of atmospheric retrievals from both \pRT{} and \texttt{ATMO}. While our original objective was to measure a robust refractory-to-volatile ratio, we found that this goal was hindered by complications of thermal dissociation and degeneracies within the retrieval setup. We discuss many of the effects at work in the retrieval results, including identifying the monotonic flattening of the H$_2$O spectral features with increasing metallicity at high temperatures. In the end, we find that combining short-wavelength data with the infrared JWST data is necessary to obtain useful constraints on the planet's composition, with the most robust measurement being a super-solar [Si/C] and sub-solar [Ti/H]. We combine these results with a thorough characterization of the host star to suggest WASP-178b's atmosphere was enriched in a carbon-depleted environment during accretion, while the titanium depletion reinforces the ubiquity of cold-trapping (Ti) on the nightside in ultra-hot Jupiters. 

Overall, the opportunity to use ultra-hot Jupiters to measure chemical inventories inaccessible to all other gas-giant atmosphere remains an important goal. Refractory-to-volatile elemental abundances can break important degeneracies in planet formation models \citep[][]{chachan:2023}. The challenge is then to understand both the stellar abundances and the planetary atmospheres robustly enough to interpret such observations. For WASP-178b, this may require more data, including continuous wavelength coverage to account for effects like dissociation, phase curves to measure the planet-wide temperature profiles and transit baseline, and high-resolution observations to probe trace species, like the products of dissociation, e.g., OH. The exploration and interpretation of ultra-hot Jupiters will benefit from more modeling to understand underlying physical processes at work in ultra-hot atmospheres, as well as more complex, physically-motivated retrieval setups.
 
\section*{Acknowledgments}
This research is based on observations made with the NASA/ESA Hubble Space Telescope obtained from the Space Telescope Science Institute, which is operated by the Association of Universities for Research in Astronomy, Inc., under NASA contract NAS 5–26555. These observations are associated with programs 16086 and 16450. This work is based in part on observations made with the NASA/ESA/CSA James Webb Space Telescope. The data were obtained from the Mikulski Archive for Space Telescopes at the Space Telescope Science Institute, which is operated by the Association of Universities for Research in Astronomy, Inc., under NASA contract NAS 5-03127 for JWST. These observations are associated with program 2055. Support for program 2055 was provided by NASA through a grant from the Space Telescope Science Institute, which is operated by the Association of Universities for Research in Astronomy, Inc., under NASA contract NAS 5-03127. All {\it HST} and {\it JWST} data used in this paper can be found in MAST: \dataset[10.17909/npy2-w182]{http://dx.doi.org/10.17909/npy2-w182}. The SDSS-IV Data Release 17 \citep{apogee:2022} Apache Point Observatory Galactic Evolution Experiment \cite[APOGEE;][]{majewski:2017} spectra our open cluster elemental abundances were derived from were collected with the APOGEE spectrograph \citep{zasowski:2013, zasowski:2017, wilson:2019, beaton:2021} on the Sloan Foundation 2.5-m Telescope \citep{gunn:2006}. As part of SDSS DR17, these spectra were reduced and analyzed with the APOGEE Stellar Parameter and Chemical Abundance Pipeline \citep[ASPCAP;][]{allendeprieto:2006, holtzman:2015, nidever:2015, aspcap} using an $H$-band line list, MARCS model atmospheres, and model-fitting tools optimized for the APOGEE effort \citep{alvarez:1998, gustafsson:2008, hubeny:2011, plez:2012, smith:2013, smith:2021, cunha:2015, shetrone:2015, jonsson:2020}. Funding for the Sloan Digital Sky Survey IV has been provided by the Alfred P. Sloan Foundation, the U.S.  Department of Energy Office of Science, and the Participating Institutions.  SDSS-IV acknowledges support and resources from the Center for High Performance Computing at the University of Utah. The SDSS website is \url{www.sdss4.org}. SDSS-IV is managed by the Astrophysical Research Consortium for the Participating Institutions of the SDSS Collaboration including the
Brazilian Participation Group, the Carnegie Institution for Science, Carnegie Mellon University, Center for Astrophysics | Harvard \& Smithsonian, the Chilean Participation Group, the French Participation
Group, Instituto de Astrof\'isica de Canarias, The Johns Hopkins University, Kavli Institute for the Physics and Mathematics of the Universe (IPMU) / University of Tokyo, the Korean Participation Group, Lawrence Berkeley National Laboratory, Leibniz Institut f\"ur Astrophysik
Potsdam (AIP),  Max-Planck-Institut f\"ur Astronomie (MPIA Heidelberg), Max-Planck-Institut f\"ur Astrophysik (MPA Garching), Max-Planck-Institut f\"ur Extraterrestrische Physik (MPE), National Astronomical Observatories
of China, New Mexico State University, New York University, University of Notre Dame, Observat\'ario Nacional / MCTI, The Ohio State University, Pennsylvania State University, Shanghai Astronomical Observatory, United Kingdom Participation Group, Universidad Nacional Aut\'onoma de M\'exico, University of Arizona, University of Colorado Boulder, University of Oxford, University of Portsmouth, University of Utah, University of Virginia, University of Washington, University of Wisconsin, Vanderbilt
University, and Yale University.  The national facility capability for SkyMapper has been funded through ARC LIEF grant LE130100104 from the Australian Research Council, awarded to the University of Sydney, the
Australian National University, Swinburne University of Technology, the University of Queensland, the University of Western Australia, the University of Melbourne, Curtin University of Technology, Monash University and the Australian Astronomical Observatory.  SkyMapper is
owned and operated by The Australian National University's Research School of Astronomy and Astrophysics.  The survey data were processed and provided by the SkyMapper Team at ANU.  The SkyMapper node of the All-Sky Virtual Observatory (ASVO) is hosted at the National Computational
Infrastructure (NCI).  Development and support of the SkyMapper node of the ASVO has been funded in part by Astronomy Australia Limited (AAL) and the Australian Government through the Commonwealth's Education Investment
Fund (EIF) and National Collaborative Research Infrastructure Strategy (NCRIS), particularly the National eResearch Collaboration Tools and Resources (NeCTAR) and the Australian National Data Service Projects (ANDS).  This work has made use of data from the European Space
Agency (ESA) mission Gaia (\url{https://www.cosmos.esa.int/gaia}),
processed by the Gaia Data Processing and Analysis Consortium (DPAC, \url{https://www.cosmos.esa.int/web/gaia/dpac/consortium}).  Funding for the DPAC has been provided by national institutions, in particular the institutions participating in the Gaia Multilateral Agreement. This publication makes use of data products from the Two Micron All
Sky Survey, which is a joint project of the University of Massachusetts and the Infrared Processing and Analysis Center/California Institute of Technology, funded by the National Aeronautics and Space Administration
and the National Science Foundation.  This publication makes use of data products from the Wide-field Infrared Survey Explorer, which is a joint project of the University of California, Los Angeles, and the
Jet Propulsion Laboratory/California Institute of Technology, funded by the National Aeronautics and Space Administration.  This research has made use of the SIMBAD database, operated at CDS, Strasbourg, France \citep{wenger:2000}.  This research has made use of the VizieR catalog access tool, CDS, Strasbourg, France.  The original description of the VizieR service was published in \citet{ochsenbein:2000}.  This research has made use of NASA's Astrophysics Data System Bibliographic Services. Based on data obtained from the ESO Science Archive Facility with DOI: \doi{10.18727/archive/25}.

\vspace{5mm}
\facilities{HST(WFC3), JWST(NIRSpec), CTIO:2MASS, Gaia, IRSA, NEOWISE, Skymapper, Sloan (APOGEE), WISE}

\software{PandExo \citep{batalha:2017},
          astropy \citep{astropy:2018},  
          PHOENIX \citep{hauschildt:1999, barman:2001}, 
          emcee \citep{foreman-mackey:2012},
          lmfit \citep{newville:2016},
          PyMultiNest \citep{buckner:2014},
          MultiNest \citep{fer08,fer09,fer19},
          \texttt{FIREFLy} \citep{rustamkulov:2022},
          \texttt{transitspectroscopy} \citep{espinoza:2022},
            \texttt{ATMO},        \citep{drummond:2016,tremblin:2017,amundsen:2014},
            \pRT{} \citep{molliere:2019,nasedkin:2024},
            \texttt{isochrones} \citep{mor15},
          \texttt{R} \citep{r24}, 
          \texttt{TOPCAT} \citep{taylor:2005}
          }
\clearpage

\appendix

\section{2D JWST/NIRSpec Spectroscopic Light Curves} \label{appendix:2DLC}

\begin{figure*}[h]
    \centering
    \includegraphics[width=1\linewidth]{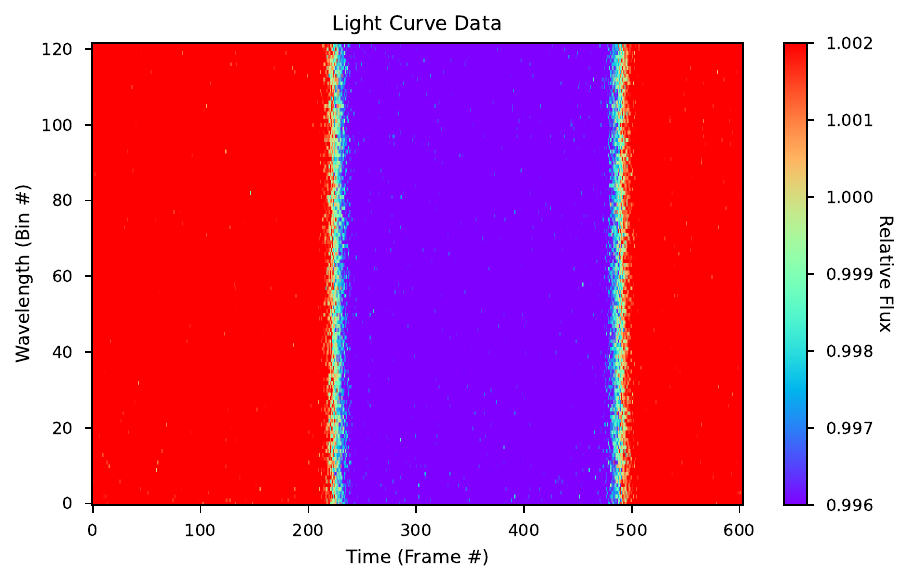}
    \caption{Two-dimensional spectroscopic light curves from JWST/NIRSpec/G395H/NRS1.}
    \label{fig:NRS1_2D_LC}
\end{figure*}

\begin{figure*}[h]
    \centering
    \includegraphics[width=1\linewidth]{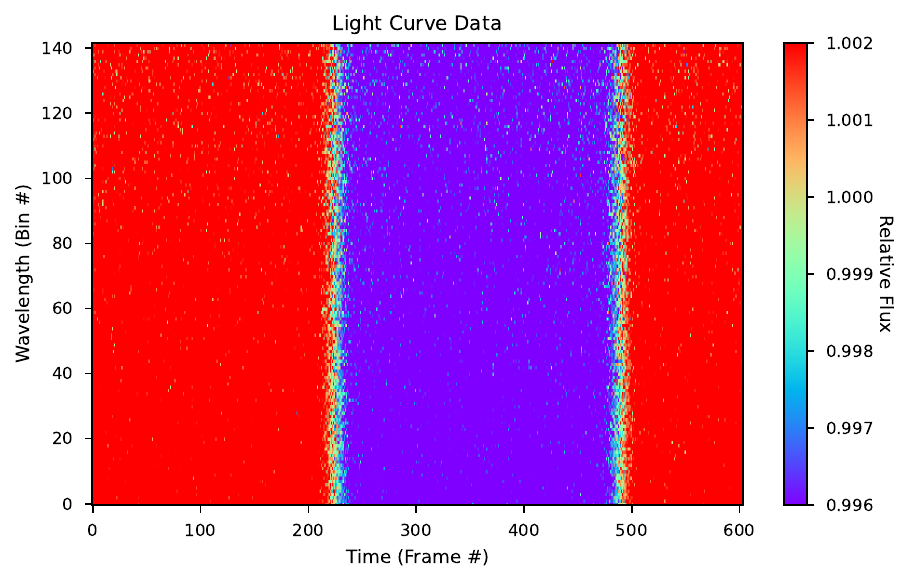}
    \caption{Two-dimensional spectroscopic light curves from JWST/NIRSpec/G395H/NRS2.}
    \label{fig:NRS2_2D_LC}
\end{figure*}

\clearpage

\section{Comparing JWST/NIRSpec Light Curve Fitting Assumptions} \label{appendix:LC}

\begin{figure*}[h]
    \centering
    \includegraphics[width=1\linewidth]{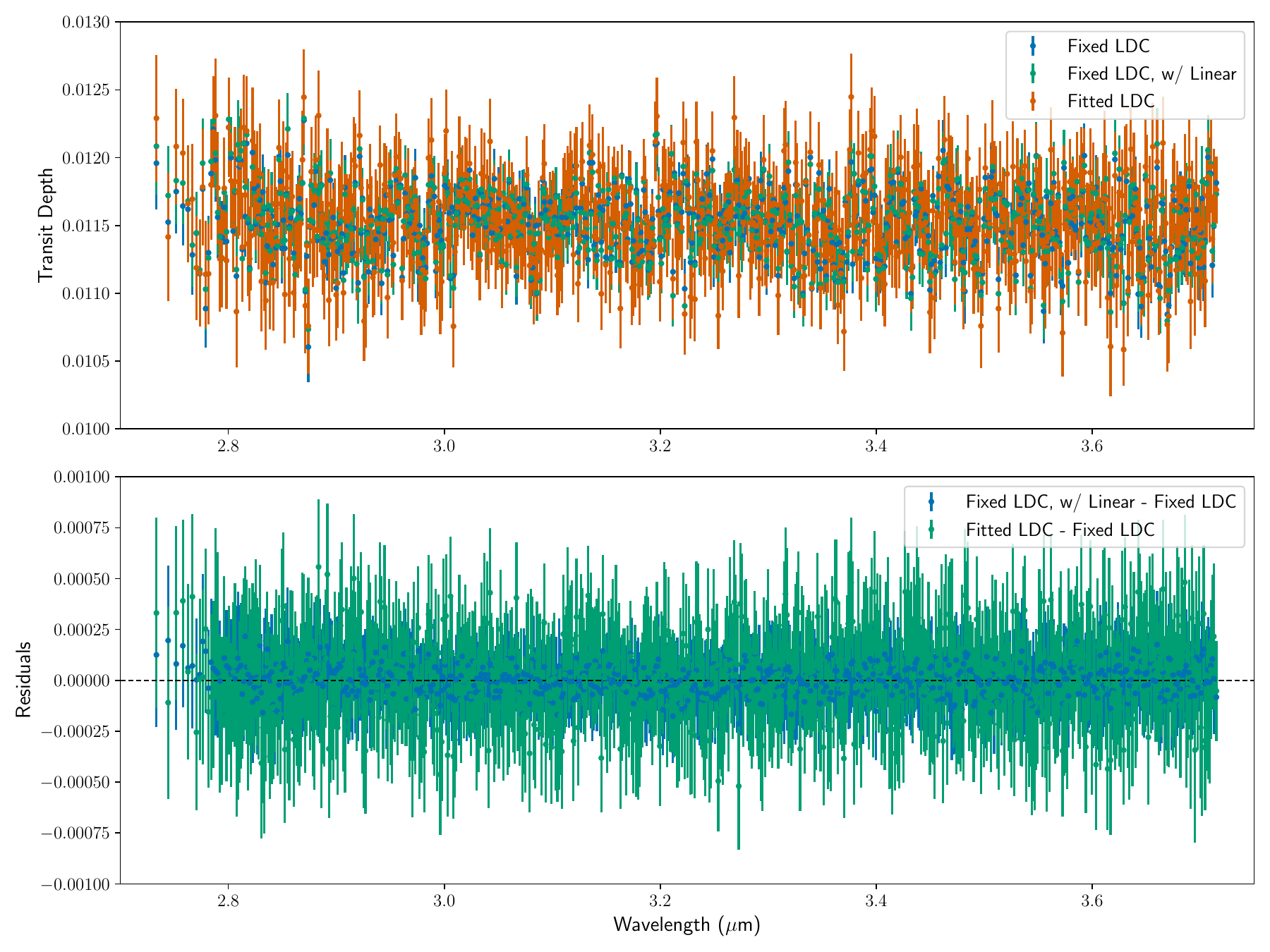}
    \caption{A comparison of different light-curve fitting methods to the NRS1 data:  1) Fixing the limb-darkening coefficients (LDCs, blue), 2) fixed LDCs and with a linear slope in time, 3) fitted LDCs. All three approaches to fitting the light curve result in data points that are very consistent within the measurement uncertainty of each point, with no discernible systematic trends.}
    \label{fig:NRS1_LC_fit}
\end{figure*}

\begin{figure*}[h]
    \centering
    \includegraphics[width=1\linewidth]{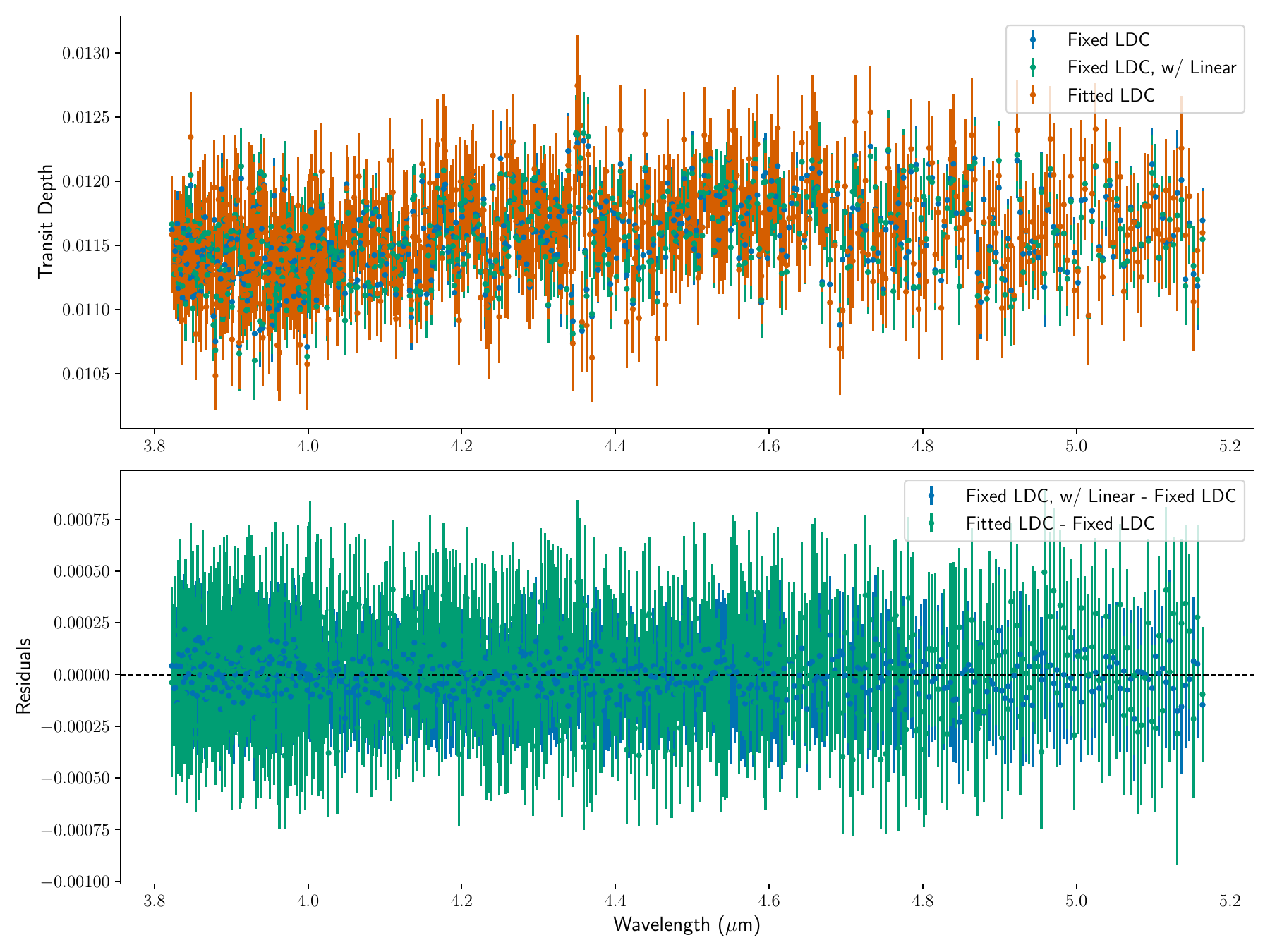}
    \caption{Same as Figure~\ref{fig:NRS1_LC_fit}, but for NRS2. As with NRS1, all three approaches to fitting the light curve result in data points that are very consistent within the measurement uncertainty of each point, with no discernible systematic trends.}
    \label{fig:NRS2_LC_fit}
\end{figure*}

\clearpage

\section{Free-Chemistry Posteriors Distributions} \label{appendix:corners}

\begin{figure*}[h]
    \centering
    \includegraphics[width=1\linewidth]{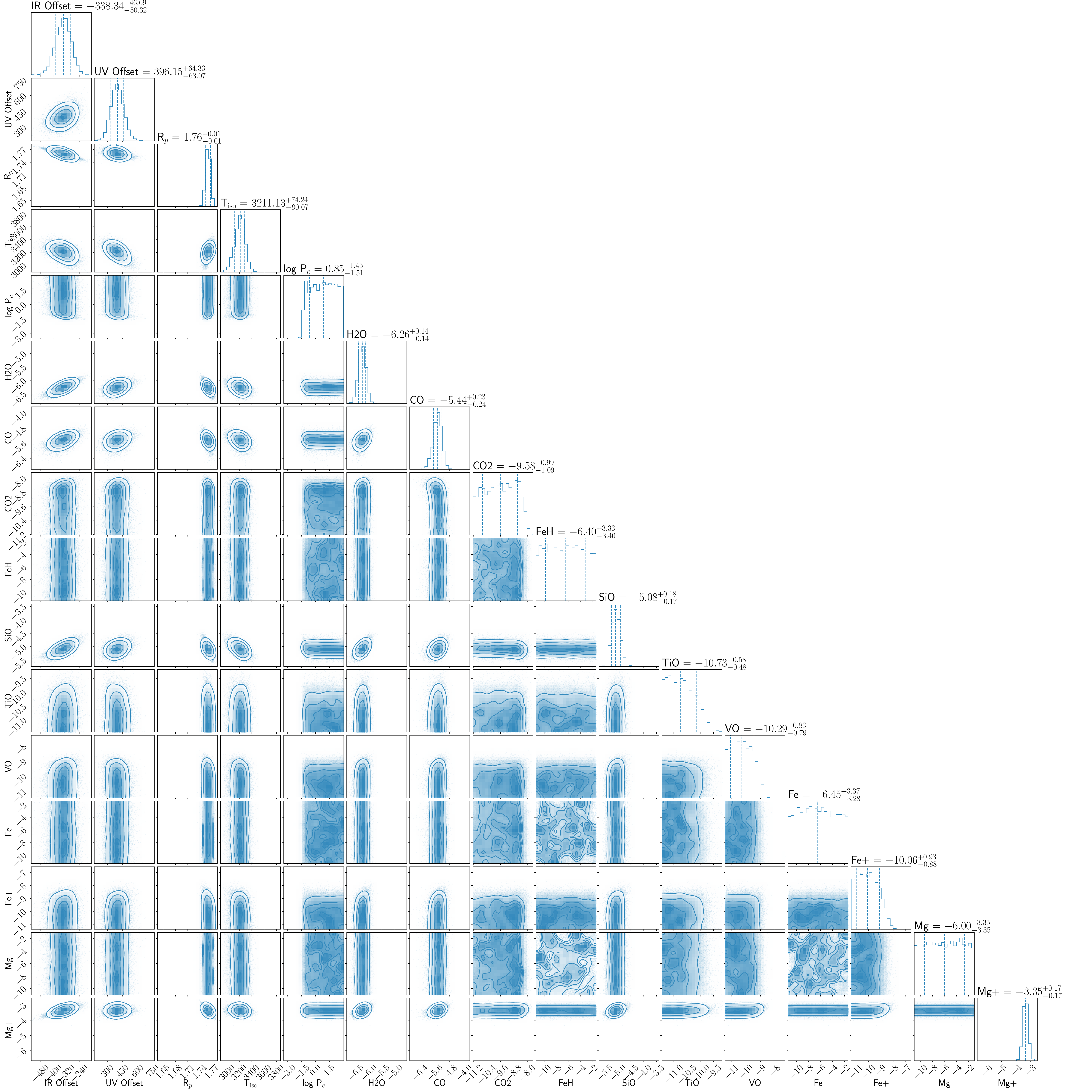}
    \caption{Posterior distribution for the \pRT{} retrieval on the full 0.2-5.1~$\mu$m spectrum, using free chemistry and an isothermal atmosphere.}
    \label{fig:full_free_corner}
\end{figure*}

\begin{figure*}[h]
    \centering
    \includegraphics[width=1\linewidth]{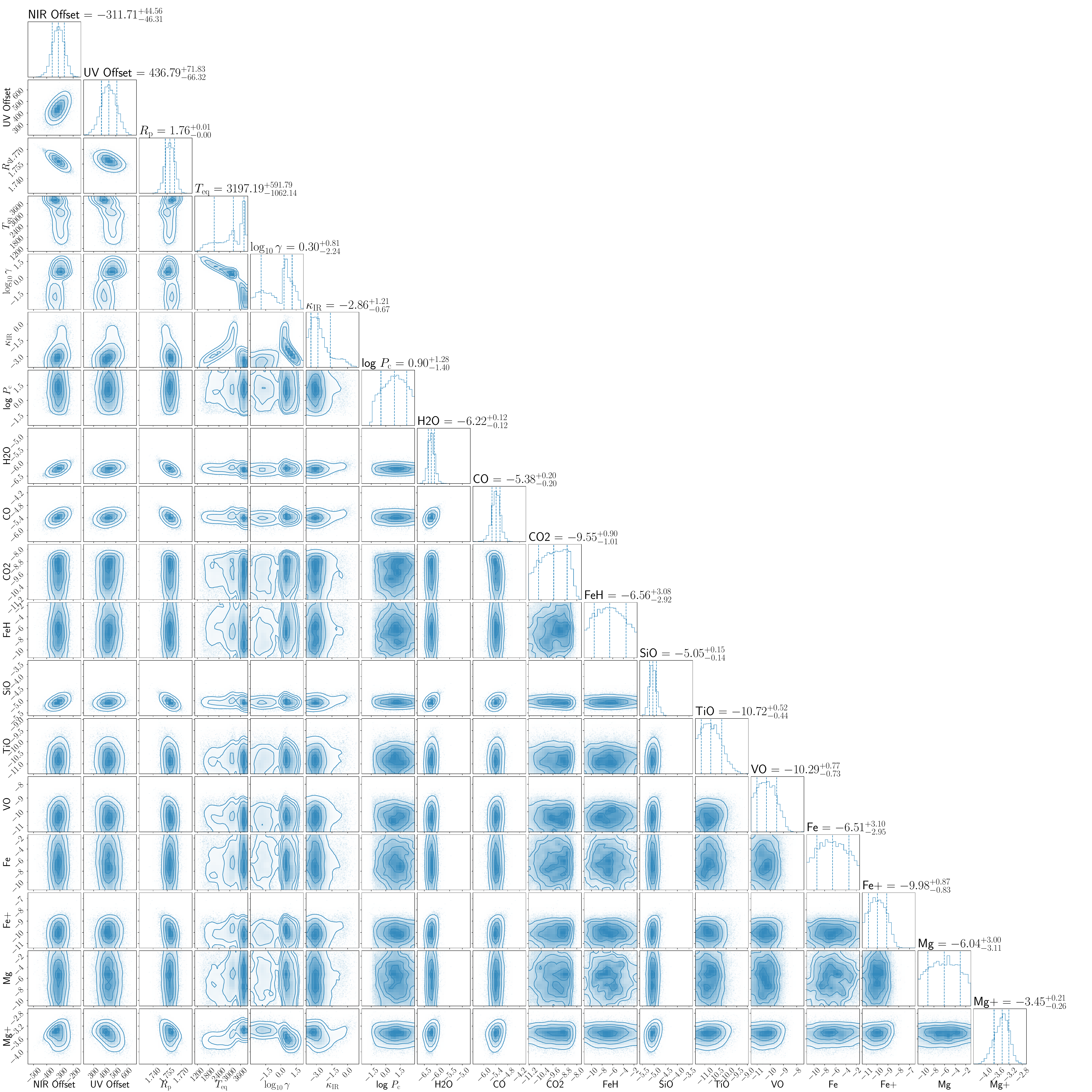}
    \caption{Posterior distribution for the \pRT{} retrieval on the full 0.2-5.1~$\mu$m spectrum, using free chemistry and the 3-parameter TP parametrization from \cite{guillot:2010}.}
    \label{fig:full_free_corner_PG}
\end{figure*}

\clearpage

\bibliographystyle{aasjournal}


\end{document}